%% file: root.tex
\documentclass[journal]{IEEEtran}
\ifCLASSINFOpdf
\usepackage[pdftex]{graphicx}
\else
\usepackage[dvips]{graphicx}
\fi
\usepackage{amsmath}
\usepackage{amssymb}
\usepackage{float}
\usepackage{subcaption}
\usepackage{multirow}	
\usepackage{etoolbox} 

\usepackage{tikz}
\usetikzlibrary{fit,shapes.misc,calc}

\input{sec/00_myCustomCommands}

\newtoggle{doublecolumn}

\begin{document}

\toggletrue{doublecolumn}

%
\title{
Performance guarantees for model-based\\Approximate Dynamic Programming\\in continuous spaces
}

%
%

\author{Paul~N.~Beuchat$^1$,
	Angelos~Georghiou$^2$,
	and~John~Lygeros$^1$,~\IEEEmembership{Fellow,~IEEE}
\thanks{
	$^1$ Automatic Control Laboratory at ETH Z\"{u}rich, Switzerland, {\tt\footnotesize \{beuchatp,jlygeros\}@ethz.ch}%
}
\thanks{
	$^2$ Desautels Faculty of Management, McGill University, Montreal,
	Canada, {\tt\footnotesize angelos.georghiou@mcgill.ca}%
}
}
\maketitle

\input{sec/01_abstract}

\input{sec/02_intro}

\input{sec/04_DP_formulation}

\input{sec/05_ADP_formulation}

\input{sec/07_bounds}

\input{sec/06_unify}

\input{sec/08_numerical}

\input{sec/10_conclusion}

%
\appendices
%

\input{sec/11_appendix_A_LPequiv_Qform}

\input{sec/11_appendix_B_LPreformulation_propositions}

\input{sec/11_appendix_C_bound_online_perf_kernel_intro}

\input{sec/11_appendix_C_bound_online_perf_proof_Qform}

\input{sec/11_appendix_D_bound_inf_norm}

\input{sec/11_appendix_E_bound_lyap_Vform}

\input{sec/11_appendix_F_unify}

\input{sec/11_appendix_G_lyap_computations}

\input{sec/11_appendix_H_SprocedureReformulation}




\bibliographystyle{IEEEtran}
\bibliography{bibliography.bib}

\end{document}

%% file: sec/00_myCustomCommands.tex
\newtheorem{theorem}{Theorem}[section]
\newtheorem{lemma}[theorem]{Lemma}
\newtheorem{definition}[theorem]{Definition}
\newtheorem{proposition}[theorem]{Proposition}

\definecolor{light-grey}{gray}{0.80}
\definecolor{mid-grey}{gray}{0.50}
\definecolor{dark-grey}{gray}{0.25}

\newcommand{\bcol}[1]{{\color{black}#1}}

\newcommand{\kcol}[1]{{\color{black}#1}}

\newcommand{\mbb}{\mathbb}

\newcommand{\mrm}{\mathrm}
\newcommand{\mcal}{\mathcal}

\newcommand{\subjto}{\text{s.t.}}

\newcommand{\tran}{\intercal}

\newcommand{\textQ}{Q}

\newcommand{\rdim}[1]{\mathbb{R}^{#1}}
\newcommand{\sdim}[1]{\mathbb{S}^{#1}}

\newcommand{\expval}[2]{\mathbb{E}_{#1}\left[#2\right]}
\newcommand{\trace}[1]{\mathrm{tr}\left(#1\right)}

\newcommand{\diag}[1]{\mathrm{diag}\left(#1\right)}

\newcommand{\argmin}[1]{\underset{#1}{\arg\min}\,\,}

\newcommand{\setdef}[2]{\left\{{#1} \,\middle|\, #2 \right\}}

\newcommand{\intd}[1]{\mathrm{d}#1}

\newcommand{\xinX}{x\!\in\!\mathcal{X}}
\newcommand{\uinU}{u\!\in\!\mathcal{U}}
\newcommand{\spaceX}{\mathcal{X}}
\newcommand{\spaceU}{\mathcal{U}}
\newcommand{\spaceXbyU}{\mathcal{X}\times\mathcal{U}}
\newcommand{\spaceXbyUcompact}{\mathcal{X}\!\times\!\mathcal{U}}
\newcommand{\funcSpaceX}{\smash{\mcal{F}(\mcal{X})}}
\newcommand{\approxFuncSpaceX}{\smash{\hat{\mcal{F}}(\mcal{X})}}
\newcommand{\approxFuncSpaceXindex}[1]{\smash{\hat{\mcal{F}}_{#1}(\mcal{X})}}
\newcommand{\funcSpaceXU}{\smash{\mcal{F}(\mcal{X} \!\times\! \mcal{U})}}
\newcommand{\approxFuncSpaceXU}{\smash{\hat{\mcal{F}}(\mcal{X} \!\times\! \mcal{U})}}
\newcommand{\approxFuncSpaceXUindex}[1]{\smash{\hat{\mcal{F}}_{#1}(\mcal{X} \!\times\! \mcal{U})}}

\definecolor{gray10}{gray}{0.10}
\definecolor{gray20}{gray}{0.20}
\definecolor{gray30}{gray}{0.30}
\definecolor{gray40}{gray}{0.40}
\definecolor{gray50}{gray}{0.50}
\definecolor{gray60}{gray}{0.60}
\definecolor{gray70}{gray}{0.70}
\definecolor{gray80}{gray}{0.80}
\definecolor{gray90}{gray}{0.90}
\definecolorset{rgb}{}{}{myred,     0.7922,0.0000,0.1255} 
\definecolorset{rgb}{}{}{myltred,   0.9569,0.6471,0.5098} 
\definecolorset{rgb}{}{}{myblue,    0.1216,0.4706,0.7059} 
\definecolorset{rgb}{}{}{myltblue,  0.6510,0.8078,0.8902} 
\definecolorset{rgb}{}{}{mygreen,   0.2000,0.6275,0.1725} 
\definecolorset{rgb}{}{}{myltgreen, 0.6980,0.8745,0.5412} 
\definecolorset{rgb}{}{}{mypurple,  0.5961,0.3059,0.6392} 
\definecolorset{rgb}{}{}{myltpurple,0.8706,0.7961,0.8941} 
\definecolorset{rgb}{}{}{myorange,  0.9882,0.5529,0.3490} 
\definecolorset{rgb}{}{}{myyellow,  0.9961,0.8784,0.5647} 

\definecolorset{rgb}{}{}{matlabred,    0.8941,0.1020,0.1098} 
\definecolorset{rgb}{}{}{matlabblue,   0.2157,0.4941,0.7216} 
\definecolorset{rgb}{}{}{matlabpurple, 0.5961,0.3059,0.6392} 
\definecolorset{rgb}{}{}{matlaborange, 1.0000,0.4980,0.0000} 



%% file: sec/01_abstract.tex

\begin{abstract}
We study both the value function and Q-function formulation of the Linear Programming  approach to Approximate Dynamic Programming. The approach is model-based and optimizes over a restricted function space to approximate the value function or Q-function.
Working in the discrete time, continuous space setting, we provide guarantees for the fitting error and online performance of the policy.
In particular, the \emph{online performance guarantee} is obtained by analyzing an iterated version of the greedy policy, and
the \emph{fitting error guarantee} by analyzing an iterated version of the Bellman inequality.
These guarantees complement the existing bounds that appear in the literature.
The Q-function formulation offers benefits, for example, in decentralized controller design, however it can lead to computationally demanding optimization problems.
To alleviate this drawback, we provide a condition that simplifies the formulation, resulting in improved computational times.
\end{abstract}

%% file: sec/02_intro.tex

\section{Introduction} \label{sec:intro}

%
%
%
%
%
%
%
%
In 1952, Bellman proposed Dynamic Programming (DP) as a solution method for discrete time stochastic optimal control problems \cite{bellman_1952_theory}.
The solution of the Bellman equation is the optimal cost-to-go function, also called the value function, which characterizes the performance of the optimal control policy.
For continuous state and action spaces, the value function takes values in an infinite dimensional function space and the policy involves computation of a multi-variate expectation.
The continuous state, input, and disturbance spaces could be discretized to leverage the extensive literature on solving the Bellman equation for finite spaces \cite{bertsekas_2017_DP_vol1,puterman_2005_MDPs}, however this is computationally infeasible even for small problems.
As such, an extensive body of literature has proposed and studied various approximation methods for continuous space problems, including a range of model-free methods that learn the value function from interactions with an un-modelled system, \cite{sutton_1988_TDlearning,bertsekas_1995_neuroBook,reviewer_suggestion_2016_modelfree,google_2015_atari}.
In this paper we focus on the model-based method named the Linear Programming (LP) approach to Approximate Dynamic Programming (ADP) \cite{schweitzer_originalMDP} that takes advantage of model knowledge for the system dynamics, cost function, and exogenous disturbance as part of the algorithm for approximating the value function.

%
The LP approach to ADP has been formulated for value functions \cite{vanRoy_linApproxDP} and \textQ-functions \cite{vanroy_decentADP}, though much of the model-based LP approach literature focuses on value function approximation.
%
The \textQ-function has the property that the optimal control policy can be expressed without involving any of the terms that describe the model.
%
This property was exploited in \cite{vanroy_decentADP,beuchat_2016_ECC_PWMQ} in a model-based setting.
%
In particular, the work of \cite{beuchat_2016_ECC_PWMQ} provides numerical evidence that the \textQ-function approximation can provide better online performance compared to the value function approximation, while the work of \cite{vanroy_decentADP} uses the \textQ-function approximation for decentralized control design in finite spaces.
%
A wide range of model-free, data-driven ADP methods also use the \textQ-function and exploit the fact that the resulting optimal policy does not directly involve any of the terms that describe the model. There are many success stories from these model-free methods, for example \cite{watkins_1989_learningFromDelayedRewards, reviewer_suggestion_2018_Qlearning_experience_replay,meyn_2017_zap_qlearning}.
%
In this paper we provide some discussion to compare model-based and model-free approaches to ADP.


\renewcommand{\arraystretch}{1.1}
\begin{table*}[t]
	\centering
	\vspace{0.0cm}
	\captionsetup{justification=centering}
	\caption{Road map to performance guarantees for the Linear Programming approach to Approximate Dynamic Programming. The bold entries represent contributions of this paper.}
	\vspace{-0.1cm}
	\begin{tabular}{|l|l|c|c|c|c|}
		\hline
		\multirow{2}{1.4cm}{Spaces}
		& \multirow{2}{1.4cm}{Bound Type}
		& \multicolumn{2}{|c|}{Non-iterated}
		& \multicolumn{2}{|c|}{Iterated}
		\\
		\cline{3-6}
		&
		& Value functions
		& \textQ-functions
		& Value functions
		& \textQ-functions
		\\
		\hline\hline
		\multirow{3}{1.2cm}{Finite}
		&
		Online performance
		&
		{\color{gray30}\cite[Theorem 1]{vanRoy_linApproxDP}}
		&
		{\color{gray30}\cite[Theorem 1]{vanroy_decentADP}}
		&
		\textemdash
		&
		\textemdash
		\\
		\cline{2-6}
		&
		Infinity norm
		&
		{\color{gray30}\cite[Theorem 2]{vanRoy_linApproxDP}}
		&
		\textemdash
		&
		\textemdash
		&
		\textemdash
		\\
		\cline{2-6}
		&
		Lyapunov-based
		&
		{\color{gray30}\cite[Theorem 3]{vanRoy_linApproxDP}}
		&
		\textemdash
		&
		\textemdash
		&
		\textemdash
		\\
		\hline
		\hline
		\multirow{3}{1.2cm}{Continuous}
		&
		Online performance
		&
		\textbf{Theorem \ref{theorem:online_performance_bound_iterated_Vform}}
		&
		\textbf{Theorem \ref{theorem:online_performance_bound_iterated_Qform}}
		&
		\textbf{Theorem \ref{theorem:online_performance_bound_iterated_Vform}}
		&
		\textbf{Theorem \ref{theorem:online_performance_bound_iterated_Qform}}
		\\ 
		\cline{2-6}
		&
		Infinity norm
		&
		{\color{gray30}\cite[\S 4.2]{boyd_iteratedBellman}}
		&
		{\color{gray30}\cite[Theorem 4.1]{beuchat_2016_ECC_PWMQ}}
		&
		{\color{gray30}\cite[\S 4.2]{boyd_iteratedBellman}}
		&
		{\color{gray30}\cite[Theorem 4.1]{beuchat_2016_ECC_PWMQ}}
		\\
		\cline{2-6}
		&
		Lyapunov-based
		&
		\textbf{Theorem \ref{theorem:bound_lyapunov_iterated_Vform}}
		&
		\textbf{Theorem \ref{theorem:bound_lyapunov_iterated_Qform}}
		&
		\textbf{Theorem \ref{theorem:bound_lyapunov_iterated_Vform}}
		&
		\textbf{Theorem \ref{theorem:bound_lyapunov_iterated_Qform}}
		\\
		\hline
		
	\end{tabular}
	\\
	\vspace{0.1cm}
	Note: Entries marked with ~``\textemdash"~ are bounds that do not exist in the literature for the finite space setting.
	\vspace{-0.2cm}
	\label{tab:bounds_roadmap}
\end{table*}
\renewcommand{\arraystretch}{1.0}

%

%
%
%
Motivated by the empirical success of the model-based LP approach, \cite{novoa_2009_ADPforVehicleRouting,stellato_2017_ADPforPowerElectronics}, a key challenge is to provide theoretical guarantees on the quality of the approximation and the online performance.
%
%
In \cite{vanRoy_linApproxDP}, the authors presented a variant of the LP approach with theoretical guarantees for finite space problems.
%
They provided three guarantees for the value function formulation: (i) a bound on the \emph{online performance} of the control policy, (ii) a bound on how close the approximate value function is to the optimal in an \emph{infinity norm} sense, and (iii) a bound on how close the approximate value function is to the optimal using a \emph{Lyapunov-based} analysis.
%
%
%
A number of works use \cite{vanRoy_linApproxDP} as a basis for deriving additional performance guarantees.
%
An online performance bound for \textQ-functions was developed in \cite{vanroy_decentADP} for the finite space setting.
%
An infinity norm bound for value function approximation was provided in \cite{boyd_iteratedBellman} for the continuous space setting by considering an iterated version of the Bellman inequality.
%
An iterated version of the infinity norm bound for \textQ-functions was given in continuous spaces  by \cite{beuchat_2016_ECC_PWMQ}.
%
A Lyapunov-based bound was presented in \cite{vivekFarias_2012_smoothedLPapproach} for finite space by analyzing a smoothed version of the LP approach.
%
%
%
As many practical control problems involve continuous state, action, and disturbance spaces, it is valuable to derive online performance and Lyapunov-based bounds for the continuous space setting.
%
In this paper we address this gap in the literature for value functions and \textQ-functions.

%
The first contribution of this paper derives novel theoretical guarantees for the value function and \textQ-function approximation using the model-based LP approach in continuous spaces, while the second contribution improves the scalability of the formulation. In particular, the contributions of the paper are:
\begin{itemize}
	%
	\item %
		We prove a continuous space online performance bound by analyzing an iterated version of the greedy policy.
		When using the non-iterated greedy policy, our bounds form the counterpart to the bounds derived in \cite[Theorem 1]{vanRoy_linApproxDP} and \cite[Theorem 1]{vanroy_decentADP} for finite spaces.
	
	\item %
		We prove a continuous space Lyapunov-based bound by analyzing the iterated Bellman inequality.
		When using the non-iterated Bellman inequality, our bounds forms the counterpart to the bound derived in \cite[Theorem 3]{vanRoy_linApproxDP} for finite spaces.
		Additionally, our bounds contain \cite[\S 4.2]{boyd_iteratedBellman} and \cite[Theorem 4.1]{beuchat_2016_ECC_PWMQ} as a special case with a Lyapunov function that is constant for all states and inputs.
	
	\item %
		Approximating the \textQ-function using the LP formulation can be computationally demanding. We provide a condition that substantially decreases the optimization problem size for the \textQ-function formulation, making the method suitable for practical applications.
\end{itemize}

%
%
The existing results and contributions of this paper are summarized in Table~\ref{tab:bounds_roadmap} for the performance guarantees.
%
In support of the contributions, we provide numerical results to demonstrate the bounds, the performance of the iterated policy, and the potential of \textQ-functions for continuous space distributed control applications.
%
%
%
Section \ref{sec:dp} presents the Dynamic Programming formulation.
Section \ref{sec:adp} introduces the approximation methods and the iterated policy.
Section \ref{sec:bounds} provides the theoretical guarantees for both the value function and \textQ-function formulations and contrasts with theoretical results from the model-free literature.
Section \ref{sec:unify} provides conditions under which the \textQ-function formulation can be simplified.
Section \ref{sec:numerical} uses numerical examples to demonstrate the theory.

\emph{Notation:}
$\rdim{}_+$ ($\rdim{}_{++}$) is the space of non-negative (positive) scalars;
$\mathbb{S}_n$ is the space of symmetric matrices of size $n$;
$\mbb{N}$ is the space of positive integers;
$I_n$ is the $\smash{n\!\times\! n}$ identity matrix;
given $\smash{f\!:\!\mcal{X}\!\rightarrow\!\rdim{}}$, the infinity norm is $\smash{ \|f\|_\infty \!=\! \sup\nolimits_{x\in\mcal{X}} |f(x)| }$, and the weighted 1-norm is $\smash{ \left\|\, f \,\right\|_{1,c} \!=\! \int_{\mcal{X}} |f(x)| c(x) \intd{x} }$.

%% file: sec/04_DP_formulation.tex

\section{Dynamic Programming (DP) Formulation} \label{sec:dp}

\subsection{Problem Formulation and Assumptions} \label{sec:dp_prob_form_and_assumptions}

We consider infinite horizon, stochastic optimal control problems with a discounted cost objective. The state of the system at time $t$ is denoted by  $x_t \!\in\! \mcal{X} \subseteq \rdim{n_x}$.
%
The state is influenced by control decisions $u_t \!\in\! \mcal{U} \subseteq \rdim{n_u}$, and stochastic disturbances $\xi_t \!\in\! \Xi \subseteq \rdim{n_\xi}$ distributed according to probability measure $\mu_\xi$ that is used in all expectations.
%
The state evolves according to $\smash{x_{t+1} = g\left( x_t , u_t , \xi_t \right)}$, where $\smash{g:\mcal{X} \!\times\! \mcal{U} \!\times\! \Xi \rightarrow \mcal{X}}$.
%
At time $t$, the system incurs the stage cost $\gamma^t \, l\left(x_t,u_t\right)$, where $\gamma \!\in\! \left[0,1\right)$ is the discount factor.
By $\Pi$ we denote the set of all feasible policies, i.e., $\left\{ \pi(\cdot) \,:\, \pi(x) \in \mcal{U},\, \forall\, x\in\mcal{X} \,\right\}$, with $\pi(\cdot)$ measurable, see \cite[Definition 2.2.3]{hernandez_2012_discreteTimeMCP}.
We restrict our attention to deterministic stationary policies and define the online performance for a fixed policy and initial state $x$ as,
	\begin{equation} \label{eq:onlinePerformane}
		V_\pi(x) \,=\, \expval{}{ \sum\nolimits_{t=0}^{\infty} \, \gamma^t \, l(x_t,\pi(x_t)) \,\middle|\, x_0 = x }
	\end{equation}
The objective is to find the policy that minimizes \eqref{eq:onlinePerformane}.

To pose this problem in the DP formulation, we work in the same setting as \cite[Section 6.3]{hernandez_2012_discreteTimeMCP}, specifically under \cite[Assumptions 4.2.1(a), 4.2.1(b), 4.2.2]{hernandez_2012_discreteTimeMCP}.
%
The assumptions ensure that from the class of time-varying stochastic policies, the minimum is attained by a stationary deterministic policy, see \cite[Theorem 4.2.3]{hernandez_2012_discreteTimeMCP}.
%
Additionally, under the assumptions it can be seen that an initial state distribution $\nu$ and an admissible policy $\pi$ define a Markov chain. Let $P_{\nu}^{\pi}\left[\cdot\right]$ denote the probability distribution of the state at time $t$, given that the initial states are distributed according to $\nu$ that is concentrated on $\mcal{X}$ and the system evolves autonomously under the fixed policy $\pi$.
%
Finally, $\funcSpaceXU$ and $\funcSpaceX$ are defined as the vector spaces of bounded, real-valued, Borel-measurable functions on $\spaceXbyUcompact$ and $\spaceX$ respectively, where \cite[Definition 6.3.2, 6.3.4]{hernandez_2012_discreteTimeMCP} provides the definitions of boundedness.

\subsection{Bellman Equation and Operator} \label{sec:dp_bellman_eq}

We now re-cast the stochastic optimal control problem in the dynamic programming formulation. The value function $\smash{V^\ast:\mcal{X} \rightarrow \rdim{}}$ represents the optimal cost-to-go from any state of the system if the optimal control policy is played, and is the solution of the Bellman equation \cite{bellman_1952_theory},
	\begin{equation} \label{eq:bellman}
		\hspace{-0.0cm}
		V^\ast(x) = \underbrace{\inf_{u\in\mcal{U}} \, \overbrace{l\left( x , u \right) + \gamma \, \expval{}{V^\ast\left(g\left(x,u,\xi\right)\right)}}^{(\mcal{T}_u V^\ast)(x,u) \,=\, Q^\ast(x,u)}}_{(\mcal{T} V^\ast)\left(x\right)}
			\,,
			\;\, \forall \, \xinX
			\,.
	\end{equation}
$\mcal{T}$ is known as the Bellman operator, and the $\mcal{T}_u$ operator is used to define an auxiliary function $\smash{Q^\ast:\spaceXbyUcompact \rightarrow \rdim{}}$ that represents the cost of making decision $u$ now and then playing optimally.
%
The Bellman equation in terms of $Q^\ast$ is thus,
\begin{equation} \label{eq:bellman_Qform} 
	\begin{aligned}
		\hspace{-0.0cm}
		Q^\ast(x,u) 
		\,&=\, \underbrace{ l(x,u) + \gamma \, \expval{}{\inf_{v \in \mcal{U}} Q^\ast\left(g\left(x,u,\xi\right),v\right)} }_{\left( F Q^\ast \right)(x,u)}
			,
	\end{aligned}
\end{equation}
for all $\xinX$ and all $\uinU$. The $F$-operator is the equivalent of $\mcal{T}$ for the so-called \textQ-functions.
The \textQ-function is an example of a \emph{post-decision value function} \cite[\S 4.6]{powell_ADPBook}.

The optimal policy can be defined using $V^\ast$ or $Q^\ast$ by,
\begin{subequations} \label{eq:bellmanpolicy}
	\begin{align}
		\pi^\ast(x)
			\,=&\, \argmin{u\in\mcal{U}} \, l\left( x , u \right) \, + \, \gamma \, \expval{}{V^{\ast}\left(g\left(x,u,\xi\right)\right)}
			\,,
			\label{eq:bellmanpolicy_Vform}
		\\
		=&\, \argmin{u\in\mcal{U}} \, Q^{\ast}(x,u)
			\,.
			\label{eq:bellmanpolicy_Qform}
	\end{align}
\end{subequations}
%
Note that evaluating \eqref{eq:bellmanpolicy_Vform} requires use of the dynamics, stage cost, and expectation with respect to $\xi$, whereas \eqref{eq:bellmanpolicy_Qform} involves only $Q^\ast$.
%
The existence of a $V^\ast$, $Q^\ast$, and $\pi^\ast$ that are Borel-measurable and attain the infimum is ensured by \cite[Assumptions 4.2.1(a), 4.2.1(b), 4.2.2]{hernandez_2012_discreteTimeMCP}.

\subsection{LP Reformulation for \textQ-functions} \label{sec:dp_lp}

Inspired by the LP reformulation of \eqref{eq:bellman} \cite{hernandez_2012_discreteTimeMCP}, we derive an LP whose optimal solution $Q^\ast$ solves equation \eqref{eq:bellman_Qform}.
%
%
%
Equation \eqref{eq:bellman_Qform} can be relaxed to an inequality,
	\begin{equation} \label{eq:F_operator_inequality} 
		\begin{aligned}
			Q(x,u) \leq F Q(x,u) \,,\quad \forall \, x \!\in\! \mcal{X} \,,\, u \!\in\! \mcal{U}
				\,,
		\end{aligned}
	\end{equation}
called the \emph{$F$-operator inequality}. One can show that operator $F$ is monotone, and satisfies value iteration convergence \cite{vanroy_decentADP}. Therefore any $Q$ satisfying \eqref{eq:F_operator_inequality} will be a point-wise under-estimator of $Q^\ast$.
%
Hence a solution of the following program,
	\begin{equation} \label{eq:LP_approach_to_DP_Qform}
	\begin{aligned}
	\hspace{-0.35cm}
		\max_{Q \in \mcal{F}(\mcal{X}\times\mcal{U})}
			\,&\quad \int_{\mcal{X}\times\mcal{U}} \, Q(x,u) \,\, c(x,u) \, \intd{x} \intd{u}
		\\
		\subjto 
			\,&\quad Q(x,u) \leq F Q(x,u) \,,\quad \forall \, x \!\in\! \mcal{X} \,,\, u \!\in\! \mcal{U}
			\,,
	\end{aligned}
	\end{equation}
coincides with the solution of \eqref{eq:bellman_Qform} for $c$-almost all ($c$-a.a) $\smash{(x,u) \!\in\! \spaceXbyUcompact}$, where $c(\cdot,\cdot)$ is a finite measure on $\spaceXbyUcompact$ that assigns positive mass to all open subsets of $\spaceXbyUcompact$; see Appendix \ref{app:LPequiv_Qform} for details.
The equivalence between \eqref{eq:bellman_Qform} and \eqref{eq:LP_approach_to_DP_Qform} requires that $\smash{\mcal{F}(\mcal{X} \!\times\! \mcal{U})}$ is the function space over which the decision variable $Q$ is optimized, see \cite[\S 6.3]{hernandez_2012_discreteTimeMCP}. Intuitively speaking, $\funcSpaceXU$ is rich enough to satisfy $\smash{Q \!\leq\! F Q}$ with equality, point-wise for all $\xinX$ and all $\uinU$.

The feasible region of \eqref{eq:LP_approach_to_DP_Qform} can be increased by using an iterated $F$-operator inequality.
%
A \textQ-function satisfying $\smash{Q\leq F^M Q}$, with $M \!\in\! \mathbb{N}$, will be a point-wise under-estimator of $Q^\ast$.
%
By $F^M$ we denote $M$ applications of the $F$ operator, and under \cite[Assumptions 4.2.1(a), 4.2.1(b), 4.2.2]{hernandez_2012_discreteTimeMCP} we have that $\smash{F^M Q \xrightarrow{M\to\infty} Q^\ast}$.
%
The same reasoning as with \eqref{eq:LP_approach_to_DP_Qform} also establishes that a solution of the following program:
	\begin{equation} \label{eq:LP_approach_to_DP_iterated_Qform}
		\begin{aligned}
			\hspace{-0.35cm}
			\max_{Q \in \mcal{F}(\mcal{X}\times\mcal{U})}
				\,&\quad \int_{\mcal{X}\times\mcal{U}} \, Q(x,u) \,\, c(x,u) \, \intd{x} \intd{u}
			\\
			\subjto
				\,&\quad Q(x,u) \leq F^\bcol{M} Q(x,u) \,,\quad \forall \, x \!\in\! \mcal{X} \,,\, u \!\in\! \mcal{U}
				\,,
		\end{aligned}
	\end{equation}
coincides with the solution of \eqref{eq:bellman_Qform} for $c$-a.a $\smash{(x,u) \!\in\! \spaceXbyUcompact}$.

The constraint in \eqref{eq:LP_approach_to_DP_iterated_Qform} is non-linear in $Q$ due to nested minimizations and expectations. A linear reformulation is obtained by introducing additional decision variables and constraints. Following the lines of \cite[Theorem 2]{vanroy_decentADP} and \cite[\S 3.4]{boyd_iteratedBellman}, problem \eqref{eq:LP_approach_to_DP_iterated_Qform} is equivalent to the infinite dimensional linear program:
%
	\begin{equation} \label{eq:LP_approach_to_DP_iterated_Qform_full}
	\begin{aligned}
		\hspace{-0.20cm}
		\max_{Q_j,V_j}
			\,&\quad \int_{\mcal{X} \times \mcal{U}} \, Q_0(x,u) \,\, c(x,u) \, \intd{x} \intd{u}
		\\
		\hspace{-0.20cm}
		\subjto
			\,&\quad \smash{ Q_j \in \funcSpaceXU }
			,\,
			\smash{ V_j \in \funcSpaceX }
			,
			&& \hspace{-0.15cm} \smash{j \!=\! 0,\dots,M\!-\!1}
			,  \hspace{-0.20cm}
		\\
		\,&\quad Q_{j}(x,u) \,\leq\, \mcal{T}_u V_{j}(x,u)
			\,,
			&& \hspace{-0.15cm} \smash{j \!=\! 0,\dots,M\!-\!1}
			,  \hspace{-0.20cm}
		\\
		\,&\quad V_{j}(x) \,\leq\, Q_{j+1}(x,u)
			\,,
			&& \hspace{-0.15cm} \smash{j \!=\! 0,\dots,M\!-\!2}
			,  \hspace{-0.20cm}
		\\
		\,&\quad V_{M-1}(x) \,\leq\, Q_{0}(x,u)
			\,,
	\end{aligned}
	\end{equation}
where the inequality constraints hold for all $\xinX$ and $\uinU$.
%
The propositions necessary to show the equivalence between \eqref{eq:LP_approach_to_DP_iterated_Qform} and \eqref{eq:LP_approach_to_DP_iterated_Qform_full} are given in Appendix \ref{app:LPreformulation_propositions}.

The introduction of the iterated $F$-operator inequality is seemingly unnecessary as it does not change the solution of problem \eqref{eq:LP_approach_to_DP_Qform}, however, it can improve the approximation quality in Section \ref{sec:adp} where the decision variables are restricted to a finite dimensional space.
%
The LP reformulation \eqref{eq:LP_approach_to_DP_iterated_Qform_full} is necessary for applying the approximation techniques of Section \ref{sec:adp}.

\vspace{0.1cm}
\subsection{Sources of Intractability} \label{sec:dp_intractabilities}

Solving \eqref{eq:LP_approach_to_DP_iterated_Qform_full} for $Q^\ast$, and implementing \eqref{eq:bellmanpolicy_Qform}, is in general intractable. The difficulties can be categorized as:
	\begin{enumerate}
		\renewcommand{\labelenumi}{(D\theenumi)}
		\item $\funcSpaceX$ and $\funcSpaceXU$ are infinite dimensional spaces;
		
		\item Problem \eqref{eq:LP_approach_to_DP_iterated_Qform_full} has infinite number of constraints;
		
		\item Objective of \eqref{eq:LP_approach_to_DP_iterated_Qform_full} involves a multidimensional integral;
		
		\item The $\mcal{T}_u$-operator involves an infinite dimensional integral over $\xi$;
		
		\item Since $Q^\ast$ can be any element of $\funcSpaceXU$, the policy \eqref{eq:bellmanpolicy_Qform} may be intractable;
	\end{enumerate}

Difficulties (D1-D5) relate to the so-called  \emph{curse of dimensionality} \cite{powell_knowAboutADP}, and apply also to the iterated value function formulation in the continuous space setting \cite{boyd_iteratedBellman}.

%% file: sec/05_ADP_formulation.tex

\section{Approximate Dynamic Programming (ADP)}\label{sec:adp}

\begin{table*} [t] 
	\centering
	\caption{Examples of overcoming (D2-D5)}
	\vspace{-0.1cm}
	\begin{tabular}{|c|l|l|l|}
		\hline
		Ref.		& Problem instance studied:   & Class of basis functions  & Overcome (D2) by:
		\\
		\hline
		\cite{boyd_iteratedBellman}, \cite{boyd_2011_minmax}, \cite{boyd_2013_iteratedApproxValueFunctions}
		&
		Linear-quadratic problems
		&
		Quadratic
		&
		S-procedure
		\\
		\hline
		\cite{lasserre_2009_soc_via_occupation_measures}, \cite{boyd_iteratedBellman}, \cite{summers_ADPwithSOS}
		&
		Polynomial problems
		&
		Polynomial
		&
		Sum-of-squares
		\\ 
		\hline
		\cite{vanroy_sampDP}, \cite{boyd_2012_quadraticADP}, \cite{sutter_2014_ADPsamp}
		&
		Finite, linear-quadratic, non-linear
		&
		Finite, quadratic, non-linear
		&
		Sampling
		\\
		\hline
		\cite{nikos_2013_ADPforReachability}, \cite{nikos_2015_ADPforReachability_arXiv}
		&
		Stochastic reachability
		&
		Radial basis functions
		&
		Sampling
		\\
		\hline
		\cite{swaroop_2010_pwcValueFunc}
		\cite{swaroop_2011_pwcValueFunc_forPerimeterPatrol}
		&
		Perimeter surveillance
		&
		Piecewise-constant
		&
		Exact Reformulation
		\\
		\hline
		
	\end{tabular}
	\vspace{-0.2cm}
	\label{tab:D2_D5_reformuation_summary}
\end{table*}

\subsection{The Approximate LP} \label{sec:adp_approxlp}

As suggested in \cite{schweitzer_originalMDP}, we restrict the value functions and \textQ-functions to take values in the span of a finite family of basis functions $\hat{V}_j^{(i)} \!:\! \mcal{X} \!\rightarrow\! \rdim{}$ and $\hat{Q}_j^{(i)} \!:\! \spaceXbyU \!\rightarrow\! \rdim{}$. We parameterize the restricted function spaces as
\begin{equation} \label{eq:approx_func_spaces}
	\begin{aligned}
		\approxFuncSpaceXindex{j} =& \setdef{\sum\nolimits_{i=1}^{K} \, \alpha_j^{(i)} \, \hat{V}_j^{(i)}(x)}{ \alpha_j^{(i)}\in\rdim{},\,}
			,
		\\
		\hspace{-0.1cm}
		\approxFuncSpaceXUindex{j} =& \setdef{\sum\nolimits_{i=1}^{K} \beta_j^{(i)} \hat{Q}_j^{(i)}\smash{(x,u)}}{\beta_j^{(i)}\in\rdim{}}
			,
	\end{aligned}
\end{equation}
for $j\!=\!0,\dots,M$. The subscript $j$ is used to highlight that the restricted function space can be different for each of the value functions and \textQ-functions. If desired, all of the restricted spaces can be taken to be the same.

An approximate solution of  \eqref{eq:LP_approach_to_DP_iterated_Qform} is obtained by the program:
	\begin{equation} \label{eq:LP_approach_to_ADP_iterated_Qform}
		\begin{aligned}
			\hspace{-0.35cm}
			\max_{\hat{Q} \in \approxFuncSpaceXUindex{0}}
				\,&\quad \int_{\mcal{X}\times\mcal{U}} \, \hat{Q}(x,u) \,\, c(x,u) \, \intd{x} \intd{u}
			\\
			\subjto
				\,&\quad \hat{Q}(x,u) \leq F^{\bcol{M}} \hat{Q}(x,u) \,,\quad \forall \, \xinX \,,\, \uinU
				\,,
		\end{aligned}
	\end{equation}
where the only change from \eqref{eq:LP_approach_to_DP_iterated_Qform} was to replace $\funcSpaceXU$ by $\approxFuncSpaceXUindex{0}$. The optimization variables are now the $\beta_0^{(i)}$'s in the definition of $\approxFuncSpaceXUindex{0}$.
%
To apply existing methods for the LP approach to ADP, we make the constraint in \eqref{eq:LP_approach_to_ADP_iterated_Qform} linear by applying  Proposition \ref{proposition:F_operator_inequality_reformulation} and \ref{proposition:F_operator_iterated_inequality_reformulation} with all the additional value functions and \textQ-functions restricted to $\approxFuncSpaceXindex{j}$ and $\approxFuncSpaceXUindex{j}$ respectively.
%
The additional decision variables and constraints introduced by this linear reformulation are a drawback that we address in Section \ref{sec:unify}.

In general, a solution of \eqref{eq:LP_approach_to_ADP_iterated_Qform}, denoted $\hat{Q}^\ast$, will not solve the Bellman equation \eqref{eq:bellman_Qform}.
%
The following lemma, which follows from \cite[Lemma 1]{vanRoy_linApproxDP}, provides the intuition that $\hat{Q}^\ast$ is the closest under-estimator of $Q^\ast$ weighted by $c(\cdot,\cdot)$.

\vspace{0.1cm}

\begin{lemma} \label{lemma:approxLP_for_Q_solves_min_1norm}
	$\hat{Q}$ is an optimal solution of \eqref{eq:LP_approach_to_ADP_iterated_Qform} if and only if it is an optimal solution of the following program
		\begin{equation} \label{eq:LP_approach_to_ADP_iterated_Qform_1norm}
		\begin{aligned}
			\hspace{-0.35cm}
			\min_{\hat{Q} \in \approxFuncSpaceXUindex{0}}
				\,&\quad \left\| \, Q^\ast \,-\, \hat{Q} \, \right\|_{1,c(x,u)}
			\\
			\subjto
				\,&\quad \text{same as \eqref{eq:LP_approach_to_ADP_iterated_Qform}}
				\,.
		\end{aligned}
	\end{equation}
\end{lemma}

A natural choice for the online policy is to replace $Q^\ast$ in equation \eqref{eq:bellmanpolicy} with the solution of \eqref{eq:LP_approach_to_ADP_iterated_Qform},
	\begin{equation} \label{eq:approxpolicy_Qform}
		\begin{aligned}
			\hat{\pi}(x)
				\,=\, \arg\min_{u\in\mcal{U}} \, \hat{Q}^\ast(x,u)
				\,,
		\end{aligned}
	\end{equation}
often referred to as the \emph{greedy policy}.
%
%
A good approximation of the optimal \textQ-function is one for which the online performance of \eqref{eq:approxpolicy_Qform} is near optimal. Although Lemma \ref{lemma:approxLP_for_Q_solves_min_1norm} shows that $\hat{Q}^\ast$ is the closest approximate \textQ-function for a given set of basis functions, it reveals nothing about the sub-optimality of policy \eqref{eq:approxpolicy_Qform}. In Section \ref{sec:bounds} we show that the online performance of \eqref{eq:approxpolicy_Qform} can be bounded by how well $\hat{Q}^\ast$ approximates $Q^\ast$.

Problem \eqref{eq:LP_approach_to_ADP_iterated_Qform} overcomes difficulty (D1) as $\approxFuncSpaceXindex{j}$ and $\approxFuncSpaceXUindex{j}$ are parameterized by a finite dimensional decision variable.
%
There are a number of choices of $\approxFuncSpaceXindex{j}$ and $\approxFuncSpaceXUindex{j}$ that address (D2-D5). The possible choices depend on the class of the stage cost and dynamics, the description of $\spaceX$ and $\spaceU$, and the distribution of the exogenous disturbance.
%
Table \ref{tab:D2_D5_reformuation_summary} summarizes examples found in the literature, where the applicability, approximation quality, and computational burden depends on the problem data and design choices made when a practitioner implements the chosen algorithm.

In \eqref{eq:LP_approach_to_DP_iterated_Qform_full} the specific choice of $c(\cdot,\cdot)$ does not affect the optimal solution. This is no longer the case in \eqref{eq:LP_approach_to_ADP_iterated_Qform} where the choice of $c(\cdot,\cdot)$ plays a central role in determining the quality of $\hat{Q}^\ast$.
%
Lemma \ref{lemma:approxLP_for_Q_solves_min_1norm} suggests that one can influence the approximation quality by an appropriate choice of $c(\cdot,\cdot)$, which is commonly referred to as the \emph{relevance weighting}.
%
To partly alleviate the dependency on the choice of relevance weighting, \cite{beuchat_2016_ECC_PWMQ} suggests solving \eqref{eq:LP_approach_to_ADP_iterated_Qform} for multiple choices of $c\smash{(\cdot,\cdot)}$, and using the point-wise maximum from the family of approximations in the greedy policy. They argue that improved online performance can be achieved with this approach.
%
Note that if the restricted function space is chosen such that $\smash{Q^\ast \!\in\! \approxFuncSpaceXUindex{0}}$, then the optimal solution of \eqref{eq:LP_approach_to_ADP_iterated_Qform} is $Q^\ast$ as long as $c(\cdot,\cdot)$ assigns positive mass to all open subsets of $\spaceXbyU$.

For completeness and comparison, we state without derivation that approximate iterated LP for the value function formulation of ADP, as introduced in \cite{boyd_iteratedBellman},
\begin{equation} \label{eq:LP_approach_to_ADP_iterated_Vform}
	\begin{aligned}
		\hspace{-0.35cm}
			\max_{\hat{V} \in \approxFuncSpaceXindex{0}}
			\,&\quad \int_{\mcal{X}} \, \hat{V}(x) \,\, c(x) \, \intd{x}
		\\
		\subjto
			\,&\quad \hat{V}(x) \leq \mcal{T}^{M} \hat{V}(x) \,,\quad \forall \, \xinX
			\,.
	\end{aligned}
\end{equation}
Weighting $c(\cdot)$ here is the counterpart of the relevance weighting in the objective of \eqref{eq:LP_approach_to_ADP_iterated_Qform} and similarly it plays a central role in determining the quality of $\hat{V}^\ast$.
The constraint is called the iterated Bellman inequality and the LP reformulation of the non-linear operator $\mcal{T}^M$ is given in \cite[\S 3.4]{boyd_iteratedBellman}.

We note that, under the assumptions of Section \ref{sec:dp_prob_form_and_assumptions}, programs \eqref{eq:LP_approach_to_ADP_iterated_Qform} and \eqref{eq:LP_approach_to_ADP_iterated_Vform} are always feasible.
Specifically, under \cite[Assumption 4.2.1(a)]{hernandez_2012_discreteTimeMCP} that the stage cost is non-negative, the choice $\smash{\alpha_j^{(i)}=\beta_j^{(i)}=0}$ for all $i$, $j$, is feasible for both the iterated $F$-operator and Bellman inequality constraints.

\subsection{Iterated Greedy Policy}  \label{sec:adp_improvedPolicy}

The following policy attempts to bridge the gap between finite horizon and two stage problems. Given $\smash{ D \!\in\! \mbb{N} \!\cup\! \{0\} }$ and an approximate \textQ-function, we define the iterated greedy policy by
	\begin{equation} \label{eq:approxpolicy_iterated_Qform}
	\begin{aligned}
		\hat{\pi}(x)
			\,=\, \arg\min_{u\in\mcal{U}} \, F^D \hat{Q}(x,u)
			\,.
	\end{aligned}
	\end{equation}
The policy may improve upon \eqref{eq:approxpolicy_Qform} for any $D \!\geq\! 1$, where we use the convention that $F^0 Q = Q$ and hence \eqref{eq:approxpolicy_Qform} and \eqref{eq:approxpolicy_iterated_Qform} coincide when $\smash{D\!=\!0}$.
%
However, computing this iterated policy is complicated by the nested expectations and minimizations arising from the $F^D \hat{Q}$ term. Using similar arguments, an iterated greedy policy using an approximate value function, $\hat{V}$, would be,
	\begin{equation} \label{eq:approxpolicy_iterated_Vform}
		\begin{aligned}
			\hspace{-0.10cm}
			\hat{\pi}(x)
			=
			\argmin{u\in\mcal{U}} \, l\left( x , u \right) + \gamma \, \expval{}{ \left( \mcal{T}^D \hat{V} \right) \left(g\left(x,u,\xi\right)\right)}
				,
		\end{aligned}
	\end{equation}
which also involves nested expectations and minimizations, and coincides with the usual greedy policy for $\smash{D\!=\!0}$.

Writing out the iterations of the $F$ or $\mcal{T}$ operator, it can be seen that the iterated greedy policy is exactly the generic form of a $D$-stage stochastic programming problem \cite[section 3.1]{shapiro_2014_lectures_on_stoch_prog}.
%
Popular approximate solution methods for such stochastic programs are Model Predictive Control (MPC) \cite{rawlings_1999_MPCtextbook,camacho_2007_MPC} and Affine Decision Rules (ADR) \cite{ben-tal_adjustable_2004,angelos_2010_generalizedDR}.
%
In \cite{morari_2014_MPCBook} the authors analyze and provide algorithms for computing an MPC policy parametric in the current state $x$, referred to as \emph{explicit MPC}. For example, when the dynamics are linear and the stage cost quadratic, as is the case in the Section \ref{sec:numerical} examples, the explicit MPC policy is shown to be piecewise linear \cite[\S 6.3]{morari_2014_MPCBook}.
%
Solving  \eqref{eq:approxpolicy_iterated_Qform} or \eqref{eq:approxpolicy_iterated_Vform} with an MPC approach would be equivalent to a finite horizon MPC formulation, with a time horizon of $D$ steps, and $\hat{Q}$ or $\hat{V}$ as the terminal cost.
%
Further details on the connection between ADP and MPC policies are given in \cite{bertsekas_2005_fromADPtoMPC}.

In Section \ref{sec:bounds}, we give a bound on the sub-optimality of the online performance achieved by \eqref{eq:approxpolicy_iterated_Qform} or \eqref{eq:approxpolicy_iterated_Vform}. This indicates that a tighter performance bound can be achieved through the iterated greedy policy. In Section \ref{sec:numerical} we use a numerical example to demonstrate the potential of this interpretation.

%% file: sec/07_bounds.tex

\section{PERFORMANCE BOUNDS FOR ADP} \label{sec:bounds}

In this section, we present performance guarantees for the continuous space setting. The \emph{online performance} bounds in Section \ref{sec:bounds_online} and \emph{Lyapunov-based} bounds in Section \ref{sec:bounds_fitting_lyap} are novel for the continuous space setting and represent a contribution of this paper.
%
To assist the reader, Table \ref{tab:bounds_roadmap} summarizes the proposed bounds and those found in the literature, \cite{vanRoy_linApproxDP}, \cite{vanroy_decentADP}, \cite{beuchat_2016_ECC_PWMQ}, and \cite{boyd_iteratedBellman}.

Note that the bounds in Section \ref{sec:bounds_fitting_infty} and \ref{sec:bounds_fitting_lyap} require that the restricted functions spaces \eqref{eq:approx_func_spaces} are all the same, which we denote as $\approxFuncSpaceX$ and $\approxFuncSpaceXU$ throughout this section.

\subsection{Online Performance Bound}  \label{sec:bounds_online}

We present first a bound on the online performance of playing the iterated greedy policy \eqref{eq:approxpolicy_iterated_Qform} or \eqref{eq:approxpolicy_iterated_Vform}.
%
These bounds only require the approximate value function or \textQ-function to be a point-wise under-estimator of $V^\ast$ or $Q^\ast$ respectively.
%
To this end, we introduce two measures: the \emph{expected state-action frequency}, $\mu$ defined on $\spaceXbyUcompact$, and its marginal on the state space, $\tilde{\mu}$ defined on $\mcal{X}$, called the \emph{expected state frequency}. For any Borel sets $\smash{\Gamma \in \mcal{B}(\mcal{X} \!\times\! \mcal{U})}$ and $\smash{B \in \mcal{B}(\mcal{X})}$ the measures are defined as:
	\begin{equation} \label{eq:definition_occupancy_measures}
		\begin{aligned}
			\mu(\Gamma) \,:=&\, \sum\nolimits_{t=0}^{\infty} \, \gamma^t \, P_{\nu}^{\pi} \left[ \left( x_t , \pi(x_t) \right) \in \Gamma \right]
				\,,
			\\
			\tilde{\mu}(B) \,:=&\, \mu(B \times \mcal{U}) \,=\, \sum\nolimits_{t=0}^{\infty} \, \gamma^t \, P_{\nu}^{\pi} \left[ x_t \in B \right]
				\,,
		\end{aligned}
	\end{equation}
See \cite[6.3.6]{hernandez_2012_discreteTimeMCP} for further details.
%
One can show that $(1-\gamma) \, \tilde{\mu}$ is a probability measure.
%
From Section \ref{sec:dp_prob_form_and_assumptions} it is clear that $V_\pi$ in \eqref{eq:onlinePerformane} is a point-wise over-estimator of $V^\ast$.
Given a function $\smash{Q : \spaceXbyU \rightarrow \rdim{}}$, define the following: $Q|_{\pi}(x) := \smash{Q(x,\pi(x))}$.

\vspace{0.2cm}

\begin{theorem} \label{theorem:online_performance_bound_iterated_Qform}
	Let $\hat{Q} : \spaceXbyU \rightarrow \rdim{}$ be such that $\hat{Q}(x,u) \leq Q^\ast(x,u)$ for all $\xinX$ and all $\uinU$, and let $\hat{\pi} : \mcal{X} \rightarrow \mcal{U}$ be a $D$-iterated policy defined in \eqref{eq:approxpolicy_iterated_Qform}. Then the sub-optimality of the online performance is bounded as,
	\begin{equation} \nonumber
		\begin{aligned}
			\hspace{-0.1cm}
			\left\| \, V_{\hat{\pi}} - V^\ast \, \right\|_{1,\nu}
			\,\leq\,
			\frac{1}{1-\gamma} \, \left\|\, \left.Q^\ast\right|_{\hat{\pi}} - \left.\left( F^{D} \hat{Q} \right)\right|_{\hat{\pi}} \,\right\|_{1,(1-\gamma)\tilde{\mu}}
				\,.
		\end{aligned}
	\end{equation}
\end{theorem}
\vspace{0.1cm}
The proof is given in Appendix \ref{app:proof_bound_online_performance}.

\vspace{0.2cm}

\begin{figure}[t]
	\centering
	\includegraphics[width=0.24\textwidth]{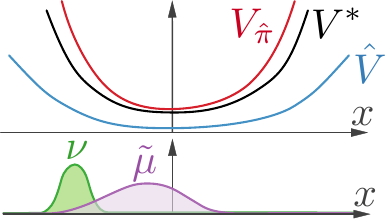}
	\caption[Visualizing the components of the online performance bound]
	{
		The upper plot shows the online performance of policy $\hat{\pi}$ is a point-wise over-estimator of $V^\ast$, and that by assumption of Theorem \ref{theorem:online_performance_bound_iterated_Vform} $\hat{V}$ is a point-wise under-estimator. The lower plot highlights that the 1-norm weightings in Theorems \ref{theorem:online_performance_bound_iterated_Qform} and \ref{theorem:online_performance_bound_iterated_Vform}, $\nu$ and $\tilde{\mu}$, can differ significantly.
	}
	\label{fig:online_bound_visualisation}
\end{figure}

\begin{theorem} \label{theorem:online_performance_bound_iterated_Vform}
	Let $\hat{V} : \mcal{X} \!\rightarrow\! \rdim{}$ be such that $\hat{V}(x) \leq V^\ast(x)$ for all $\xinX$, and let $\hat{\pi} : \mcal{X} \!\rightarrow\! \mcal{U}$ be a $D$-iterated policy defined in \eqref{eq:approxpolicy_iterated_Vform}. Then the sub-optimality of the online performance is bounded as,
	\begin{equation} \nonumber
		\begin{aligned}
			\left\| \, V_{\hat{\pi}} - V^\ast \, \right\|_{1,\nu} \,\leq\, \frac{1}{1-\gamma} \,\, \left\|\, V^\ast - \left( \mcal{T}^{D} \hat{V} \right) \,\right\|_{1,(1-\gamma)\tilde{\mu}}
				\,.
		\end{aligned}
	\end{equation}
\end{theorem}
\vspace{0.1cm}
The proof is a minor adaptation of the proof of Theorem \ref{theorem:online_performance_bound_iterated_Qform}.
%
%
Notice that for $\smash{D\!=\!0}$, Theorems \ref{theorem:online_performance_bound_iterated_Qform} and \ref{theorem:online_performance_bound_iterated_Vform} are reminiscent of the finite space versions, \cite[Theorem 1]{vanroy_decentADP} and \cite[Theorem 1]{vanRoy_linApproxDP} respectively. The proofs, however, require a different analysis due to the consideration of continuous spaces.
%
Fig. \ref{fig:online_bound_visualisation} visualizes the quantities involved.

The following insights apply to Theorem \ref{theorem:online_performance_bound_iterated_Qform} and \ref{theorem:online_performance_bound_iterated_Vform}:
\begin{itemize}
	\item They provide the reassurance for continuous space problems that when policy \eqref{eq:approxpolicy_iterated_Qform} or \eqref{eq:approxpolicy_iterated_Vform} uses an under-estimator, the sub-optimality of the online performance is bounded by how closely $\hat{Q}$ or $\hat{V}$ fits $Q^\ast$ or $V^\ast$ respectively.
	
	\item They motivate the potential benefit of considering a $D$-iterated policy based on an under-estimator. Although $F$ and $\mcal{T}$ are not contractive with respect to the weighted 1-norm, it is expected that the right hand side gets smaller as $D$ increases, and hence the online sub-optimality is more tightly bounded.
\end{itemize}

\vspace{0.2cm}

\subsection{Infinity-norm Bound}  \label{sec:bounds_fitting_infty}

We present now a result that bounds the fitting of $\hat{Q}^\ast$ or $\hat{V}^\ast$ relative to $Q^\ast$ or $V^\ast$, by how close $Q^\ast$ or $V^\ast$ is to the span of the basis functions.
%
These bound were reported in \cite{beuchat_2016_ECC_PWMQ} and \cite{boyd_iteratedBellman} and are included here for completeness.

\vspace{0.1cm}

\begin{theorem} \label{theorem:inf_norm_bound_iterated_Qform}
	Let $Q^\ast$ be the solution of \eqref{eq:bellman_Qform} and $\hat{Q}^\ast$ be the solution of \eqref{eq:LP_approach_to_ADP_iterated_Qform} for a given choice $\approxFuncSpaceXU$ and $c(\cdot,\cdot)$ then,
	\begin{equation} \label{eq:theorem:inf_norm_bound_iterated_Qform_bound}
		\begin{aligned}
			\left\| Q^\ast - \hat{Q}^\ast \right\|_{1,c(x,u)} \,\leq\, \frac{2}{1-\gamma^M} \,\, \inf _{\hat{Q}\in\approxFuncSpaceXU} \, \left\| Q^\ast - \hat{Q} \right\|_\infty
				.
		\end{aligned}
	\end{equation}
	
\end{theorem}
\vspace{-0.1cm}
The proof was first reported in our preliminary study \cite[Theorem 4.1]{beuchat_2016_ECC_PWMQ}. It is included in Appendix \ref{app:proof_bound_inf_norm_Qform} in the interest of completeness.

\vspace{0.1cm}

\begin{theorem} \label{theorem:inf_norm_bound_iterated_Vform}
	Let $V^\ast$ be the solution of \eqref{eq:bellman} and $\hat{V}^\ast$ be the solution of \eqref{eq:LP_approach_to_ADP_iterated_Vform} for a given choice $\approxFuncSpaceX$ and $c(\cdot)$ then,
	\begin{equation} \nonumber
		\begin{aligned}
			\left\| \, V^\ast \,-\, \hat{V}^\ast \, \right\|_{1,c(x)} \,\leq\, \frac{2}{1-\gamma^M} \,\, \inf_{\hat{V} \in \approxFuncSpaceX} \, \| V^\ast - \hat{V} \|_\infty
				.
		\end{aligned}
	\end{equation}
\end{theorem}
\vspace{0.1cm}
The proof is given in \cite[\S 4.3]{boyd_iteratedBellman}.

\vspace{0.2cm}

Comparing the left-hand-side in Theorem \ref{theorem:inf_norm_bound_iterated_Vform} to the right-hand-side in Theorem \ref{theorem:online_performance_bound_iterated_Vform}, the choice $c(\cdot)\!=\!(1\!-\!\gamma)\tilde{\mu}(\cdot)$ means that the online performance is bounded by Theorem \ref{theorem:inf_norm_bound_iterated_Vform}. For Theorems \ref{theorem:inf_norm_bound_iterated_Qform} and \ref{theorem:online_performance_bound_iterated_Qform} to be combined in a similar way, the relevance weighting $c(\cdot,\cdot)$ should satisfy,
	\begin{equation} \nonumber
		\begin{aligned}
			\left\|\, \left.Q^\ast\right|_{\hat{\pi}} - \left.\left( F^{D} \hat{Q} \right)\right|_{\hat{\pi}} \,\right\|_{1,\tilde{\mu}}
				\,=\,
				\left\| Q^\ast - \hat{Q}^\ast \right\|_{1,c(x,u)}
				\,.
		\end{aligned}
	\end{equation}
In both cases, choosing $c$ as described is a difficult task since it is a circular requirement: the choice of $c$ affects the solution of the approximate LP, which in turn affects the approximate policy, which affects the expected state frequency $\tilde{\mu}$, which in turn affects the desired relevance weighting $c$.

The following insights apply to Theorem \ref{theorem:inf_norm_bound_iterated_Qform} and \ref{theorem:inf_norm_bound_iterated_Vform}:
\begin{itemize}
	\item As $\smash{\| Q^\ast - \hat{Q} \|_\infty}$ and $\smash{\| V^\ast - \hat{V} \|_\infty}$ may be very large, the bounds may be too conservative for practical use. We investigate this limitation, which affects all similar bounds in literature, through numerical examples in Section \ref{sec:numerical}.
	
	\item The right-hand-side of the bounds hold for any choice of the relevance weightings. Thus, the bounds do not provide any intuition for how to choose $c(\cdot,\cdot)$ or $c(\cdot)$.
	
	\item A large $M$ tightens the bound via the $\gamma^M$ term. The upper bound on $M$ is dictated by the size of the approximate LP that can be solved in the time frame available.
	
	\item The right-hand-side of the bounds may be infinite in some cases. Consider for example a linear-quadratic problem on unbounded spaces where $V^\ast$ is known to be quadratic. If $\approxFuncSpaceX$ is the space of affine functions, then $\| V^\ast - \hat{V} \|_\infty$ is infinite for all elements from $\approxFuncSpaceX$.
\end{itemize}

\subsection{Lyapunov-based Bound}  \label{sec:bounds_fitting_lyap}

Finally, we derive Lyapunov-based bounds that are novel for the continuous space setting.
%
To this end, for functions $\smash{V : \mcal{X} \rightarrow \rdim{}}$ we define an operator $H_{V}$,
\begin{equation} \nonumber
	\begin{aligned}
		\left( H_{V} V \right)(x) \,=\, \max\nolimits_{u\in\mcal{U}} \expval{}{ V\left( f(x,u,\xi) \right) }
			\,,
	\end{aligned}
\end{equation}
and for functions $\smash{Q:\spaceXbyU \rightarrow \rdim{}}$ we define an operator $H_{Q}$,
\begin{equation} \nonumber
	\begin{aligned}
		\left( H_{Q} Q \right)(x,u) \,=\, \max\nolimits_{v\in\mcal{U}} \expval{}{ Q\left( f(x,u,\xi) , v \right) }
			\,.
	\end{aligned}
\end{equation}
Given that the system is in state $x$, the function $(H_{V} V)(x)$ represents the worst case expected value of $V$ at the next state. For \textQ-functions, given further that action $u$ will be applied, the function $(H_{Q} Q)(x,u)$ represents the worst case expected value two times steps into the future.
It is readily shown that both $H_{V}$ and $H_{Q}$ are monotone operators.

Given function $V$ and $Q$, let,
\begin{equation} \nonumber
	\begin{aligned}
		\beta_V \,=&\, \max\limits_{\quad\,\,\,\, x\in\mcal{X} \quad\,\,\,\,} \, && \gamma \, (H_{V} V)(x) \,/\, V(x)
			\,,
		\\
		\beta_Q \,=&\, \max\limits_{(x,u)\in(\mcal{X} \times \mcal{U})} \, && \gamma \, (H_{Q} Q)(x,u) \,/\, Q(x,u)
			\,,
	\end{aligned}
\end{equation}
be the maximum ratio of the worse case expected value at a future time step, to the value in the current state(-by-input).

\vspace{0.1cm}

\begin{definition} \label{def:lyap_func_Vform}
	A function $V: \mcal{X} \rightarrow \rdim{}_{++}$ is called a \underline{\emph{Lyapunov function}} if $\beta_V < 1$.
\end{definition}

\vspace{0.1cm}

\begin{definition} \label{def:lyap_func_Qform}
	A function $Q: \spaceXbyU \rightarrow \rdim{}_{++}$ is called a \underline{\emph{Lyapunov \textQ-function}} if $\beta_Q < 1$.
\end{definition}

\vspace{0.1cm}

For any positive function, $\smash{V:\mcal{X} \rightarrow \rdim{}_{++}}$, let $1/V$ denote the map $x \mapsto 1/V(x)$, and similarly for a strictly positive $\smash{Q:\spaceXbyU \rightarrow \rdim{}_{++}}$. Now we can state the bounds.

\vspace{0.1cm}

\begin{theorem} \label{theorem:bound_lyapunov_iterated_Vform}
	Let $V^\ast$ be the solution of \eqref{eq:bellman} and $\hat{V}^\ast$ be the solution of \eqref{eq:LP_approach_to_ADP_iterated_Vform} for a given choice $\approxFuncSpaceX$ and $c(\cdot)$. Then, for any Lyapunov function $\hat{V}^{+}(x) \in \approxFuncSpaceX$,
	\begin{equation} \nonumber
	\begin{aligned}
		\left\| V^\ast - \hat{V}^\ast \right\|_{1,c(x)}
			\,\leq\,
			\frac{ 2 \, \left\| \hat{V}^{+} \right\|_{1,c(x)}}{1-\beta^M_{\hat{V}^{+}}} \, \inf_{\hat{V}\in\hat{\mcal{F}}} \, \left\| V^\ast - \hat{V} \right\|_{\infty , 1/\hat{V}^{+}}
			.
	\end{aligned}
	\end{equation}
\end{theorem}
\vspace{0.1cm}
The proof is given in Appendix \ref{app:proof_bound_lyap_Vform}.

\vspace{0.1cm}

\begin{theorem} \label{theorem:bound_lyapunov_iterated_Qform}
	Let $Q^\ast$ be the solution of \eqref{eq:bellman_Qform} and $\hat{Q}^\ast$ be the solution of \eqref{eq:LP_approach_to_ADP_iterated_Qform} for a given choice $\approxFuncSpaceXU$ and $c(\cdot,\cdot)$. Then, for any Lyapunov \textQ-function $\smash{\hat{Q}^{+}(x,u) \in \approxFuncSpaceXU}$:
	\begin{equation} \nonumber
	\begin{aligned}
		\hspace{-0.2cm}
		\left\| Q^\ast - \hat{Q}^\ast \right\|_{1,c(x,u)}
			\leq
			\frac{ 2 \, \left\| \hat{Q}^{+} \right\|_{1,c(x,u)}}{1-\beta^M_{\hat{Q}^{+}}} \, \inf_{\hat{Q}\in\hat{\mcal{F}}} \, \left\| Q^\ast - \hat{Q} \right\|_{\infty , 1/\hat{Q}^{+}}
	\end{aligned}
	\end{equation}
\end{theorem}
\vspace{0.1cm}
The proof follows by modifying the proof for Theorem \ref{theorem:bound_lyapunov_iterated_Vform}.
%
When $\smash{M\!=\!1}$, Theorem \ref{theorem:bound_lyapunov_iterated_Vform} is reminiscent of the finite space version \cite[Theorem 3]{vanRoy_linApproxDP}. The proof requires an adapted analysis due to the consideration of the iterated Bellman inequality.

The following insights apply to Theorem \ref{theorem:bound_lyapunov_iterated_Vform} and \ref{theorem:bound_lyapunov_iterated_Qform}:
\begin{itemize}
	\item Theorems \ref{theorem:inf_norm_bound_iterated_Qform} and \ref{theorem:inf_norm_bound_iterated_Vform} are a special case of \ref{theorem:bound_lyapunov_iterated_Qform} and \ref{theorem:bound_lyapunov_iterated_Vform} because the function that returns a constant value for all $\xinX$, $\uinU$ is a Lyapunov function with $\beta_V=\beta_Q=\gamma$.
	
	\item As the inverse of the Lyapunov function weights the infinity norm term, the bounds may be tighter than Theorems \ref{theorem:inf_norm_bound_iterated_Qform} and \ref{theorem:inf_norm_bound_iterated_Vform}. To see this, consider that in regions where $V^\ast$ or $Q^\ast$ are large, the Lyapunov function may also be chosen to be large and hence reduce the worst case error in those regions. Section \ref{sec:numerical_1d} provides an example where, for larger $M$, the bound tightening is significant.
	
	\item The relevance weighting now appears on the right-hand side of the bound. This indicates that an appropriate choice of relevance weighting is the one which gives the tightest bound. However, finding the combination of a relevance weighting and Lyapunov function that yields the tightest bound is, in general, a difficult problem.
\end{itemize}

We refer to \cite[\S 5]{vanRoy_linApproxDP} for some discussion on the choice of Lyapunov functions for finite space problems.

\subsection{Comparison to Temporal Difference and \textQ-learning} \label{sec:unify:qlearning_comparison}

%
Model-free approaches, such as Temporal Difference (TD) learning \cite{sutton_1988_TDlearning} and \textQ-learning \cite{watkins_1989_learningFromDelayedRewards}, aim to optimize the control policy based only on data collected through interactions with the system.
%
Recent results demonstrate many successes and great potential of these methods, see for example \cite{reviewer_suggestion_2016_modelfree,google_2015_atari,reviewer_suggestion_2018_Qlearning_experience_replay,meyn_2017_zap_qlearning,reviewer_suggestion_2017_policy_gradient,google_2017_alphago}.
%
By contrast, model-based approaches assume complete and accurate knowledge of the underlying system model when synthesizing a control policy, and through this provide theoretical analysis and performance guarantees.
%
In particular, the LP approach to ADP ensures that the approximate value functions and \textQ-functions are point wise under-estimators of $V^\ast$ and $Q^\ast$, thus facilitating the theoretical guarantees presented in Sections \ref{sec:bounds_online}, \ref{sec:bounds_fitting_infty}, \ref{sec:bounds_fitting_lyap}.
%
In this section we compare aspects of the model-based LP approach to ADP with examples from the TD and \textQ-learning literature.
%
We refer the reader to \cite[Chapter 8]{sutton_2018_rlbook} for further discussion on comparing model-based and model-free approaches.

Similar to the LP approach, many variants of TD learning also use a linearly parameterized function space for the approximation architecture, for example,  \cite{sutton_1988_TDlearning,bradtke_1996_lstd,boyan_2002_lstd}. In \cite{tsitsiklis_1997_TD_analysis} the authors provide theoretical guarantees on the approximation quality of the solution from TD learning. In particular, \cite[Theorem 1]{tsitsiklis_1997_TD_analysis} provides an approximation quality bound that is reminiscent of Theorem \ref{theorem:inf_norm_bound_iterated_Vform}. However, as TD learning is designed for autonomous systems, there is no notion of an online performance guarantee.
%
For a controlled system, the Actor-Critic algorithm in \cite{konda_2003_actorcritic_journal} uses TD learning in the critic step, and in the actor step it makes gradient updates in the control policy space. The authors show convergence of their Actor-Critic algorithm to a local minimum with respect to the parametrization of the control policy. However, they do not provide any bound on the sub-optimality of the resulting policy.
%
By contrast, the model-based LP approach to ADP allows one to compute online performance guarantees such as those offered by Theorems \ref{theorem:online_performance_bound_iterated_Qform} and \ref{theorem:online_performance_bound_iterated_Vform}.

%
Recent \textQ-learning methods utilize Neural Networks for the restricted function space \cite[Section 6.3.1]{bertsekas_2017_DP_vol1}, \cite{reviewer_suggestion_2017_policy_gradient}, and demonstrate many successes, for example playing games \cite{google_2015_atari,google_2017_alphago} and regulating a two-degree-of-freedom helicopter \cite{reviewer_suggestion_2018_helicopter}.
%
This suggests that Neural Networks can be an interesting choice of restricted function space for the LP approach to ADP presented in this paper.
%
However, the non-linear nature of Neural Networks will complicate the analysis of the LP approach and makes for an interesting future research direction, potentially providing guarantees for a fixed Neural Network architecture.

%% file: sec/06_unify.tex

\section{PARTICULAR \textQ-FUNCTION FORMULATIONS} \label{sec:unify}

In this section, we consider cases for which the \textQ-function formulation can be simplified. We first present the condition which facilitates this simplification, thus making the formulation computationally efficient. We then provide two problem classes for which the condition is satisfied. In particular, this formulation can be beneficial for the decentralized control designs that we discuss in Section \ref{sec:unify_structured_functions}.

\subsection{Condition for equivalence} \label{sec:unify_lemma}

Applying Propositions \ref{proposition:F_operator_inequality_reformulation} and \ref{proposition:F_operator_iterated_inequality_reformulation} to \eqref{eq:LP_approach_to_ADP_iterated_Qform}, the approximate LP for the \textQ-function formulation is,
%
	\begin{subequations} \label{eq:PropOfEquiv_forQ}
		\begin{align}
			\hspace{-0.3cm}
			\max_{\hat{Q}_j,\hat{V}_j}
				\,&\quad \int_{\mcal{X} \times \mcal{U}} \, \hat{Q}_0(x,u) \,\, c(x,u) \, \intd{x} \intd{u}
				\nonumber
			\\
			\subjto \,&\quad \hat{Q}_j \in \approxFuncSpaceXUindex{j}
				,\,
				\hat{V}_j \in \approxFuncSpaceXindex{j}
				\,,
				&& \hspace{-0.10cm} j=0,\dots,M\!-\!1
				\,,
				\nonumber
			\\
			\,&\quad \hat{Q}_{j}(x,u) \,\leq\, \mcal{T}_u \hat{V}_{j}(x,u)
				\,,
				&& \hspace{-0.65cm} j=0,\dots,M\!-\!1
				\,,
				\label{eq:PropOfEquiv_forQ_01}
			\\
			\,&\quad \hat{V}_{j}(x) \,\leq\, \hat{Q}_{j+1}(x,u)
				\,,
				&& \hspace{-0.65cm} j=0,\dots,M\!-\!2
				\,,
				\label{eq:PropOfEquiv_forQ_02}
			\\
			\,&\quad \hat{V}_{M-1}(x) \,\leq\, \hat{Q}_{0}(x,u)
				\,,
				\label{eq:PropOfEquiv_forQ_03}
		\end{align}
	\end{subequations}
where the inequality constraints hold for all $\smash{\xinX}$ and all $\smash{\uinU}$. Now consider the following formulation with $M\!-\!1$ fewer \textQ-functions and $M\!-\!1$ fewer infinite constraints:
	\begin{subequations} \label{eq:PropOfEquiv_forV}
		\begin{align}
			\hspace{-0.3cm}
			\max_{\hat{Q}_0 , \hat{V}_j}
				\,&\quad \int_{\mcal{X} \times \mcal{U}} \, \hat{Q}_{0}(x,u) \,\, c(x,u) \, \intd{x} \intd{u}
				\nonumber
			\\
			\subjto \,&\quad \hat{Q}_{0} \in \approxFuncSpaceXUindex{0}
				,\,
				\hat{V}_{j} \in \approxFuncSpaceXindex{j}
				\,,
				&& \hspace{-0.10cm} j=0,\dots,M\!-\!1
				\,,
				\nonumber
			\\
			\,&\quad \hat{Q}_{0}(x,u) \,\leq\, \mcal{T}_u \hat{V}_{0} (x,u)
				\,,
				\label{eq:PropOfEquiv_forV_01}
			\\
			\,&\quad \hat{V}_{j-1}(x) \,\leq\, \mcal{T}_u \hat{V}_{j}(x,u)
				\,,
				&& \hspace{-0.65cm} j=1,\dots,M\!-\!1
				\,,
				\label{eq:PropOfEquiv_forV_02}
			\\
			\,&\quad \hat{V}_{M-1}(x) \leq \hat{Q}_{0}(x,u)
				\,,
				\label{eq:PropOfEquiv_forV_03}
		\end{align}
	\end{subequations}
where the inequality constraints hold for all $\xinX$ and all $\uinU$. In Lemma \ref{lemma:Qform_Vform_equivalence} below, we provide a condition for when \eqref{eq:PropOfEquiv_forQ} and \eqref{eq:PropOfEquiv_forV} are equivalent. 

\vspace{0.1cm}

\begin{lemma} \label{lemma:Qform_Vform_equivalence}
	If the sets $\approxFuncSpaceXindex{j}$ and $\approxFuncSpaceXUindex{j}$ are chosen such that for all $\smash{\hat{V}_j\in\approxFuncSpaceXindex{j}}$ there exists a $\smash{\hat{Q}_j\in\approxFuncSpaceXUindex{j}}$ with
	\begin{equation} \nonumber
		\begin{aligned}
			&\hat{Q}_j(x,u) = \mcal{T}_u\hat{V}_j(x,u) \,,\quad \forall \, \xinX \,,\,\, \uinU
				\,,
		\end{aligned}
	\end{equation}
	for $j=1,\dots,M\!-\!1$,
	then the approximate LP \eqref{eq:PropOfEquiv_forQ} and \eqref{eq:PropOfEquiv_forV} have the same optimal value and there is a mapping between feasible and optimal solutions in both problems.
\end{lemma}
\vspace{0.1cm}
The proof is given in Appendix \ref{app:Qform_equivalence}.

\subsection{Input constrained, Linear-Quadratic control} \label{sec:unify_LQ}

In the case of linear dynamics, quadratic cost function, and control actions constrained to lie in a polytopic feasible set, then the value function and \textQ-function is known to be piece-wise quadratic \cite[Theorem 6.7]{morari_2014_MPCBook}. Hence, quadratic basis functions defined as,
%
	\begin{equation} \label{eq:quadratic_basis_function_space}
	\begin{aligned}
		\approxFuncSpaceXindex{j} =&
			\setdef{ \hat{V}(x) }{
			\begin{matrix}
				V_j(x) \!=\! x^\tran P_{j}^{} x + p_{j}^{\tran} x + s_{j}^{}
				\\
				P_{j}^{} \in \sdim{n_x} \,,\,\, p_{j}^{} \in \rdim{n_x} \,,\,\, s_{j}^{} \in \rdim{}
			\end{matrix}
		}
		\\
		\approxFuncSpaceXUindex{j} \!=\!&
			\left\{\,
			\begin{matrix}
				\hat{Q}_j(x,u) \text{  such that:}\hfill
				\\
				Q_j(x,u) = \begin{bmatrix}x \\ u \end{bmatrix}^\tran P_{j}^{Q} \begin{bmatrix}x \\ u \end{bmatrix} + p_{j}^{Q} \begin{bmatrix}x \\ u \end{bmatrix} + s_{j}^{Q}
				\\
				P_{j}^{Q} \in \sdim{n_x+n_u} \,,\,\, p_{j}^{Q} \in \rdim{n_x+n_u} \,,\,\, s_{j}^{Q} \in \rdim{}
			\end{matrix}
		\,\right\}
	\end{aligned}
	\end{equation}
are reasonable choices, see \cite[\S 6]{boyd_iteratedBellman}. The $\alpha_j^{(i)}$'s and $\beta_j^{(i)}$'s from \eqref{eq:approx_func_spaces} are the coefficients of the monomials. In this setting, for any quadratic value function, the term
	\begin{equation} \nonumber
		\begin{aligned}
			\mcal{T}_u \hat{V}(x,u) =  l(x,u) + \mbb{E} [\hat{V}(g(x,u,\xi))]
				\,,
		\end{aligned}
	\end{equation}
will be quadratic in $\smash{(x,u)}$, and requires knowledge of the first and second moments of the exogenous disturbance. As $\approxFuncSpaceXUindex{j}$ is taken to be the space of all quadratic functions in $[x^\tran , u^\tran]^\tran$, the condition of Lemma \ref{lemma:Qform_Vform_equivalence} is satisfied.

\subsection{Structured \textQ-functions for decentralized control} \label{sec:unify_structured_functions}

Consider a decentralized control problem with $N$ agents. The input for each agent, $u=[u_1^\tran,\dots,u_N^\tran]^\tran$, can only depend on a locally available portion of the state vector, $x_1,\dots,x_N$, i.e., a decentralized policy is of the form,
	\begin{equation} \nonumber
		\begin{aligned}
			u \,=\, \pi_{\mrm{Decent}}(x) \,=\,
				\begin{bmatrix}
					\pi_1(x_1)^\tran
					&
					\cdots
					&
					\pi_N(x_N)^\tran
				\end{bmatrix}^\tran
				\,.
		\end{aligned}
	\end{equation}
This framework is not readily addressed by traditional DP formulations.

As the greedy policy \eqref{eq:approxpolicy_Qform} is a constrained optimization problem, decentralized control is realized if both the objective and constraint set have the required separable structure. For the constraint $\smash{u\in\mcal{U}}$ we assume that the set is separable, i.e., $\mcal{U} = \mcal{U}_1 \times \dots \times \mcal{U}_N$. As The \textQ-function is the objective of \eqref{eq:approxpolicy_Qform}, it will be separable if the \textQ-function is a sum of per-agent \textQ-functions that only depend on $u_i$ and $x_i$. Let $\mcal{S}$ denote the set of functions with the separable structure:
	\begin{equation} \label{eq:Q_decent_structure}
		\begin{aligned}
			\mcal{S} = \setdef{\hat{Q}(\cdot,\cdot)}{\hat{Q}(x,u) \,=\, \sum\limits_{i=1}^{N} \, \hat{Q}_{i}(u_i,x_i) \,\,+\,\, q(x)}
		\end{aligned}
	\end{equation}
where $q(x)$ can be any function of the full state vector. The term $q(x)$ is allowed because it does not affect the decision made by evaluating the greedy policy. This separable \textQ-function structure is as suggested in \cite{vanroy_decentADP}.

It is necessary to enforce the structural constraint \eqref{eq:Q_decent_structure} on $\hat{Q}_0$ in both \eqref{eq:PropOfEquiv_forQ} and \eqref{eq:PropOfEquiv_forV}. The remaining \textQ-functions and value functions need not have the decentralized structure enforced.
Lemma \ref{lemma:Qform_Vform_equivalence} allows a different restricted function space for $\hat{Q}_0$, and hence can be applied to the decentralized control formulation.

The value function formulation can also be used to approximate a solution to the decentralized control problem. It requires the assumption that $l(x,u) \in \mcal{S}$, and the restriction on the approximate value function that $\mbb{E} [\hat{V}(g(x,u,\xi))] \in \mcal{S}$.

%% file: sec/08_numerical.tex

%
%
%

\section{NUMERICAL RESULTS} \label{sec:numerical}

In this section we present three numerical examples to highlight various aspects of the theory presented above. The first example numerically evaluates the performance bounds from Section \ref{sec:bounds}, the second assesses the potential of the iterated approximate policy presented in Section \ref{sec:adp}, whereas the third demonstrates using \textQ-function for a distributed control setting as per Section \ref{sec:unify_structured_functions}. The second example also provides empirical evidence that the \textQ-function formulation can achieve tighter lower bounds.

In all numerical examples we use a Linear Quadratic Regulator (LQR) as a point of comparison. This is a linear state feedback controller synthesized via the Riccati equation for a system with linear dynamics and quadratic stage cost.
%
The code to generate the results is found at \cite{beuchat_2017_ADPToolbox}.

\subsection{Evaluation of Performance Bounds} \label{sec:numerical_1d}

We use a one dimensional example from \cite{boyd_iteratedBellman} with $\smash{n_x\!=\!n_u\!=\!n_\xi \!=\! 1}$ to highlight that although the iterated value function gives an tighter lower bound of the optimal cost-to-go, it can have both worse online performance, and a worse online performance bound. The dynamics, costs, and constraints are given by,
	\begin{equation} \nonumber
		\begin{aligned}
			x_{t+1} \,=&\,\, x_t - 0.5 u_t + \xi_t \,,\,\,\,&& l(x,u) \,=\, x^2 + 0.1 u^2 \,,
			\\
			\gamma \,=&\,\, 0.95\,,\,\,\,&& \;\;\;\;\;\; |u| \leq 1
		\end{aligned}
	\end{equation}
with the exogenous disturbance and initial condition distributed as $\xi_t \!\sim\! \mcal{N}(0,0.1)$ and $x_0 \!\sim\! \mcal{N}(0,\sigma_{\nu}^{2} \!=\! 10) \!=\! \nu$, respectively. The benefit of using a 1-dimensional example is that the value function and optimal policy ($V^\ast$ and $\pi^\ast$) can be effectively approximated by using a discretization method and used to directly asses the quality of approximation.

We use the space of univariate quadratics as $\approxFuncSpaceX$, without a linear term due to the problem symmetry, and we choose the state-relevance weighting as the initial state distribution, i.e.,
	\begin{equation} \nonumber
		\begin{aligned}
			\approxFuncSpaceX = \setdef{p x^2 + s}{p,s \in \rdim{}}
				\,,
				\qquad
				c(\cdot) \!=\! \nu(\cdot)
				\,.
		\end{aligned}
	\end{equation}
%
We compare approximate value functions, solved via the iterated approximate LP with $M\!=\!\{1,10,200\}$ and their respective approximate policies.

Table \ref{tab:OnlineBounds_for1DExample} shows the bound of Theorem \ref{theorem:online_performance_bound_iterated_Vform} for this example, and Table \ref{tab:FittingBounds_for1DExample} shows the bounds of Theorems \ref{theorem:inf_norm_bound_iterated_Vform} and \ref{theorem:bound_lyapunov_iterated_Vform}.
%
For completeness, details on the computation of $V^\ast$, $\hat{V}$, $V_{\hat{\pi}}$, $\mu$, and the Lyupanov functions are given in Appendix \ref{app:supplement_for_performance_bounds_numerical}.

\begin{table} [h]
	\centering
	\caption{Bounds for example \ref{sec:numerical_1d}. The last column is the percentage decrease from the right-hand-side of Theorem \ref{theorem:inf_norm_bound_iterated_Vform} to the right-hand-side of Theorem \ref{theorem:bound_lyapunov_iterated_Vform}, and $\beta_{V^+}$ is for the Lyapunov function that gives the smallest value for right-hand-side of Theorem \ref{theorem:bound_lyapunov_iterated_Vform}.}
	\begin{tabular}{|c|c c c c c|}
		\iftoggle{doublecolumn}{
			\hline
			\multirow{2}{*}{$M$} 
			& \multirow{2}{1.6cm}{LHS of Thm. \ref{theorem:inf_norm_bound_iterated_Vform} \& \ref{theorem:bound_lyapunov_iterated_Vform}}
			& \multirow{2}{1.1cm}{RHS of Thm. \ref{theorem:inf_norm_bound_iterated_Vform}}
			& \multirow{2}{1.1cm}{RHS of Thm. \ref{theorem:bound_lyapunov_iterated_Vform}}
			& \multirow{2}{*}{$\beta_{V^+}$}
			& \multirow{2}{*}{$\% \downarrow$}
		%
		}{
			\hline
			\multirow{2}{*}{$M$} 
			& \multirow{2}{1.8cm}{LHS of Thm. \ref{theorem:inf_norm_bound_iterated_Vform} \& \ref{theorem:bound_lyapunov_iterated_Vform}}
			& \multirow{2}{1.4cm}{RHS of Thm. \ref{theorem:inf_norm_bound_iterated_Vform}}
			& \multirow{2}{1.4cm}{RHS of Thm. \ref{theorem:bound_lyapunov_iterated_Vform}}
			& \multirow{2}{*}{$\beta_{V^+}$}
			& \multirow{2}{*}{$\% \downarrow$}
		%
		}
		\\
		&&&&&
		\\
		\hline
		$1$
		& $22.2$
		& $28158$
		& $27831$
		& $0.970$
		& $1.2$
		\\
		$10$
		& $16.4$
		& $3509$
		& $3161$
		& $0.972$
		& $9.9$
		\\
		$200$
		& $10.1$
		& $1408$
		& $541$
		& $0.988$
		& $61.6$
		\\
		\hline
	\end{tabular}
	\label{tab:FittingBounds_for1DExample}
\end{table}

\begin{table} [h]
	\centering
	\caption{Online performance bound of Theorem \ref{theorem:online_performance_bound_iterated_Vform} for example \ref{sec:numerical_1d}. Evaluated numerically using $10^8$ Monte Carlo simulations per controller.}
	\begin{tabular}{|c c|c c c|}
		\hline
		$M$
		& $D$
		& $\left\| V_{\hat{\pi}} \!-\! V^\ast \right\|_{1,\nu}$ 
		& $\leq$
		& $\frac{1}{1-\gamma} \, \left\| V^\ast \!-\! \left( \mcal{T}^{D} \hat{V} \right) \right\|_{1,(1\!-\!\gamma)\tilde{\mu}}$
		\\
		\hline
		\multicolumn{2}{|c|}{LQR}
		& $0.061$
		& $\leq$
		& $73.2$
		\\
		$1$ & $0$
		& $0.061$
		& $\leq$
		& $113.7$
		\\
		$10$ & $0$
		& $0.069$
		& $\leq$
		& $138.0$
		\\
		$200$ & $0$
		& $0.079$
		& $\leq$
		& $168.5$
		\\
		$200$ & $1$
		& $0.061$
		& $\leq$
		& $158.5$
		\\
		$200$ & $2$
		& $0.061$
		& $\leq$
		& $150.1$
		\\
		$200$ & $3$
		& $0.061$
		& $\leq$
		& $142.6$
		\\
		$200$ & $4$
		& $0.061$
		& $\leq$
		& $135.8$
		\\
		$200$ & $5$
		& $0.061$
		& $\leq$
		& $129.4$
		\\
		\hline
	\end{tabular}
	\label{tab:OnlineBounds_for1DExample}
\end{table}

Table \ref{tab:FittingBounds_for1DExample} shows that, for this example, the bounds of Theorems \ref{theorem:inf_norm_bound_iterated_Vform} and \ref{theorem:bound_lyapunov_iterated_Vform} are conservative, but the Lyapnov-based approach tightens the bound for all values of $M$. Interestingly, the benefit of the Lyapunov-based bound is more pronounced for larger $M$.
To understand the reasoning for this example, see that that a $\beta_{V^+}$ closer to $1$ coincides with a Lyapunov function that minimizes the term:
	\begin{equation} \nonumber
	\begin{aligned}
		\left\| \hat{V}^{+} \right\|_{1,c(x)} \, \min_{\hat{V}\in\hat{\mcal{F}}} \, \left\| V^\ast - \hat{V} \right\|_{\infty , 1/\hat{V}^{+}}
			\,.
	\end{aligned}
	\end{equation}
However, a larger $M$ is required to ensure that the denominator term $1-\beta^M_{\hat{V}^{+}}$ does not dominate the bound. For each value of $M$, there is a sweet spot that gives the tightest bound, see Appendix \ref{app:supplement_for_performance_bounds_numerical}. This highlights the benefit of deriving the Lyapunov-based bound using the iterated Bellman formulation.

As indicated by the dependence on $M$, the right-hand-side of Theorems \ref{theorem:inf_norm_bound_iterated_Vform} and \ref{theorem:bound_lyapunov_iterated_Vform} are improved by more than an order of magnitude in going from $M\!=\!1$ to $M\!=\!200$. However, as the bounds are anyway conservative, it is not clear that in general the left-hand-side of Theorem \ref{theorem:inf_norm_bound_iterated_Vform} and \ref{theorem:bound_lyapunov_iterated_Vform} should decrease as $M$ increases.
For this example, the choice $c(\cdot) \!=\! \nu(\cdot)$ means the left-hand-side is the under-estimation error of the optimal cost-to-go, $J^\ast \!=\! \int V^\ast \intd{\nu}$, and numerically agree with \cite{boyd_iteratedBellman}.

%
%
\begin{table*}[t] 
	\centering
	\caption{Results averaged over $20$ randomly generated $50$-dimensional examples for each $\gamma$. For the online performance, the expectation over $\smash{x_0\!\sim\!\nu}$ is computed using $500$ samples, and expectation with respect to $\xi$ is computed from $500$ Monte Carlo simulation each of length $2000$ time steps. For the lower bounds the expectation over $x_0$ is computed from the same $500$ samples. In order to aggregate results across different systems, the costs and computation times are normalized with respect to the average performance of the MPC controller with horizon $\smash{T\!=\!10}$. The column ``Controller computation" relates to the average computation time in milliseconds to compute the control action at each time step, using a single thread on a 3.00Ghz Xeon processor. The ratio to the controller with the highest computation load is shown in the ``speed-up" column.
	}
	
	\vspace{0.1cm}
	\begin{tabular}{|cc|l|cccc|cccc|cc|}
		\hline
		\multicolumn{3}{|c|}{\textbf{Description}} &
		\multicolumn{4}{|c|}{\textbf{Normalized cost, $\gamma = 0.95$}} &
		\multicolumn{4}{|c|}{\textbf{Normalized cost, $\gamma = 0.99$}} &
		\multicolumn{2}{|c|}{\textbf{Controller computation}}
		\\
		\multicolumn{3}{|c|}{} &
		\textbf{avg.} &
		\textbf{$\sigma$} &
		\textbf{min.} &
		\textbf{max.} &
		\textbf{avg.} &
		\textbf{$\sigma$} &
		\textbf{min.} &
		\textbf{max.} &
		\textbf{time (ms)} &
		\textbf{speed-up}
		\\
		\hline\hline
		\multirow{13}{0.0cm}{\rotatebox{90}{Online}} &
		\multirow{13}{0.1cm}{\rotatebox{90}{Performance}} &
		LQR &
		$1.1475$  &  $ 0.446$  &  $ 0.400$  &  $ 8.249$  &
		$1.1805$  &  $ 0.266$  &  $ 0.652$  &  $ 4.439$  &  $-$ & $-$
		\\
		&& $\hat{Q}^\ast$ with $M=1$, $D=0$ &
		$1.0090$  &  $ 0.325$  &  $ 0.368$  &  $ 3.322$  &
		$1.0257$  &  $ 0.187$  &  $ 0.602$  &  $ 2.611$  &  $   0.014$  &  $    80.6$
		\\
		&& $\hat{V}^\ast$ with $M=1$, $D=0$ &
		$1.0085$  &  $ 0.325$  &  $ 0.368$  &  $ 3.322$  &
		$1.0254$  &  $ 0.186$  &  $ 0.602$  &  $ 2.605$  &  $   0.014$  &  $    80.8$
		\\
		&& $\hat{V}^\ast$ with $M=50$, $D=0$ &
		$1.0044$  &  $ 0.318$  &  $ 0.374$  &  $ 3.249$  &
		$1.0110$  &  $ 0.176$  &  $ 0.607$  &  $ 2.515$  &  $   0.014$  &  $    81.4$
		\\
		&& $\hat{Q}^\ast$ with $M=50$, $D=0$ &
		$1.0043$  &  $ 0.318$  &  $ 0.375$  &  $ 3.248$  &
		$1.0109$  &  $ 0.176$  &  $ 0.608$  &  $ 2.510$  &  $   0.014$  &  $    81.4$
		\\
		&& MPC: $T=5$, $l_T=\mrm{lqr}$ &
		$1.0018$  &  $ 0.319$  &  $ 0.367$  &  $ 3.239$  &
		$1.0027$  &  $ 0.174$  &  $ 0.598$  &  $ 2.478$  &  $   0.225$  &  $     5.2$
		\\
		&& $\hat{Q}^\ast$ with $M=1$, $D=4$ &
		$1.0002$  &  $ 0.317$  &  $ 0.367$  &  $ 3.226$  &
		$1.0025$  &  $ 0.174$  &  $ 0.598$  &  $ 2.472$  &  $   0.225$  &  $     5.2$
		\\
		&& $\hat{V}^\ast$ with $M=1$, $D=4$ &
		$1.0002$  &  $ 0.317$  &  $ 0.367$  &  $ 3.226$  &
		$1.0025$  &  $ 0.174$  &  $ 0.598$  &  $ 2.470$  &  $   0.225$  &  $     5.2$
		\\
		&& $\hat{V}^\ast$ with $M=50$, $D=4$ &
		$1.0001$  &  $ 0.317$  &  $ 0.367$  &  $ 3.223$  &
		$1.0008$  &  $ 0.173$  &  $ 0.598$  &  $ 2.459$  &  $   0.225$  &  $     5.2$
		\\
		&& $\hat{Q}^\ast$ with $M=50$, $D=4$ &
		$1.0001$  &  $ 0.317$  &  $ 0.367$  &  $ 3.223$  &
		$1.0008$  &  $ 0.173$  &  $ 0.598$  &  $ 2.459$  &  $   0.225$  &  $     5.2$
		\\
		&& MPC: $T=10$, $l_T=\mrm{lqr}$ &
		$1.0000$  &  $ 0.317$  &  $ 0.367$  &  $ 3.222$  &
		$1.0000$  &  $ 0.172$  &  $ 0.598$  &  $ 2.458$  &  $   1.175$  &  $     1.0$
		\\
		\hline\hline
		&&Optimal, $V^\ast$, and $Q^\ast$   &
		\multicolumn{4}{|c|}{not available} &
		\multicolumn{4}{|c|}{not available} &
		$-$ &
		$-$
		\\
		\hline\hline
		\multirow{4}{0.0cm}{\rotatebox{90}{Lower}} &
		\multirow{4}{0.1cm}{\rotatebox{90}{Bound}} &
		$\hat{Q}^\ast$ with $M=50$ &
		$ 0.924$  &  $ 0.287$  &  $ 0.274$  &  $ 2.511$  &
		$ 0.892$  &  $ 0.139$  &  $ 0.510$  &  $ 1.924$  &     $-$  &     $-$
		\\
		&& $\hat{V}^\ast$ with $M=50$ &
		$ 0.907$  &  $ 0.275$  &  $ 0.280$  &  $ 2.342$  &
		$ 0.872$  &  $ 0.132$  &  $ 0.512$  &  $ 1.826$  &     $-$  &     $-$
		\\
		&& $\hat{Q}^\ast$ with $M=1$ &
		$ 0.853$  &  $ 0.246$  &  $ 0.274$  &  $ 2.122$  &
		$ 0.864$  &  $ 0.121$  &  $ 0.522$  &  $ 1.673$  &     $-$  &     $-$
		\\
		&& $\hat{V}^\ast$ with $M=1$ &
		$ 0.812$  &  $ 0.230$  &  $ 0.268$  &  $ 1.947$  &
		$ 0.833$  &  $ 0.113$  &  $ 0.518$  &  $ 1.516$  &     $-$  &     $-$
		\\
		\hline
	\end{tabular}
	
	\label{tab:numerical_results_NDexample}
\end{table*}

%
%

Table \ref{tab:OnlineBounds_for1DExample} shows that, for this example, the online performance bounds of Theorem \ref{theorem:online_performance_bound_iterated_Vform} is also conservative, and that the iterated policy tightens the bound.
The difficulty in choosing the state-relevance weighting is highlighted by the fact that the approximate value function with $M\!=\!200$ gives a better lower-bound of $V^\ast$ but has worse online performance.
For the $M\!=\!200$ approximate LP, it would be possible to choose a $c(x)$ different from $\nu$ that yields a value function similar to $\hat{V}^\ast$ with $M\!=\!1$.
Thus there is an inherent discrepancy between choosing a $c(x)$ that maximizes the lower-bound of $V^\ast$, useful for assessing sub-optimality, and choosing a $c(x)$ that achieves the best online performance. 

The bottom five rows of Table \ref{tab:OnlineBounds_for1DExample} show that, as expected, the iterated policy improves both the online performance and the online performance bound. As nice feature of this bound is that it theoretically converges to $0$ as $D$ increases. However, for higher dimensional systems the Bellman operator can only be approximated for a low number of iterations.
We study this in more detail in the next section.

\subsection{High-dimensional example} \label{sec:numerical_nd}

To highlight the potential of the iterated greedy policy, proposed in Section \ref{sec:adp_improvedPolicy}, on a system of higher dimension we consider an input constrained Linear Quadratic Regulator (LQR) problem. The system dynamics are $\smash{x_{t+1} = A x_{t} + B_u u + B_\xi \xi}$, with $\smash{x_{t}\!\in\!\rdim{50}}$, $\smash{u_{t}\in\rdim{6}}$, $\smash{\xi_{t}\in\rdim{50}}$, and the matrices $A$, $B_u$, $B_{\xi}$, of compatible size, describe the linear dynamics.
%
The $A$ and $B_u$ matrices are randomly generated with $A$ scaled to be marginally stable, and the results are averaged over the performance on 20 separate extractions for each $\smash{\gamma \!=\! \{0.95,0.99\}}$.
%
In all cases $B_\xi$ is an identity matrix, and the exogenous disturbance and initial condition are distributed as $\xi_t \!\sim\! \mcal{N}(0,0.1 I_{50})$ and $x_0 \!\sim\! \mcal{N}(0,9 I_{50}) \!=\! \nu$, respectively.
%
The $\spaceX$ space is unconstrained, while the $\spaceU$ space is a hyper-rectangle with the lower and upper bounds chosen to make the constrains relevant for the whole horizon.

Table \ref{tab:numerical_results_NDexample} presents the online performance results of using quadratic approximate value functions and \textQ-functions, parameterized as in \eqref{eq:quadratic_basis_function_space}. We solve \eqref{eq:PropOfEquiv_forV} with $\smash{M\!=\!1}$ and $\smash{M\!=\!50}$ and simulate both the greedy policy \eqref{eq:approxpolicy_Qform}, and the iterated greedy policy \eqref{eq:approxpolicy_iterated_Qform} and \eqref{eq:approxpolicy_iterated_Vform} with $\smash{D\!=\!4}$. As discussed in Section \ref{sec:adp_improvedPolicy}, the iterated greedy policy is approximated with an MPC-reformulation, where $\smash{D\!=\!4}$ corresponds to a prediction horizon of $\smash{T\!=\!5}$. For a comparison controller, we use MPC with prediction horizon $\smash{T\!=\!\{5,10\}}$ and the Riccati equation solution as the terminal cost function, implemented with the batch approach as detailed in \cite[\S 8.2]{morari_2014_MPCBook}.
%
The online performance is computed as $\expval{\nu}{\, \expval{\xi}{ \sum\nolimits_{t=0}^{2000} \gamma^t \, l(x_t,u_t) \middle| x_0, \hat{\pi}(\cdot) } \,}$ using $500$ Monte Carlo samples from $\nu$, and the expectation with respect to $\xi$ computed from $500$ Monte Carlo simulations each of length $2000$ time steps.
%
The table also presents lower-bounds on the value function implied by each approximation. The lower-bound is computed as $\expval{\nu}{\hat{V}(x)}$ and $\expval{\nu}{\min\nolimits_{u} \hat{Q}(x,u)}$ respectively, with the expectations computed using the same $500$ Monte Carlo samples from $\nu$.

%
For this example, $\hat{Q}^\ast$ with $\smash{M\!=\!50}$ gives a tighter lower bound when compared to $\hat{Q}^\ast$ with $\smash{M\!=\!1}$, the same trend as for the 1 dimensional example of Section \ref{sec:numerical_1d}. By contrast, the online performance of the greedy policy using $\hat{Q}^\ast$ with $\smash{M\!=\!50}$ is improved compared to using $\hat{Q}^\ast$ with $\smash{M\!=\!1}$.
%
The iterated greedy policy achieves a noticeable improvement in the online performance, compared to the standard greedy policy, both for the value function and \textQ-function formulation.

%
\begin{figure*} [t]
	\raggedright
	\begin{tikzpicture}
	\node[inner sep=0pt,anchor=south west] at (0.0cm,1.0cm){
		\includegraphics[width=5.6cm]
		{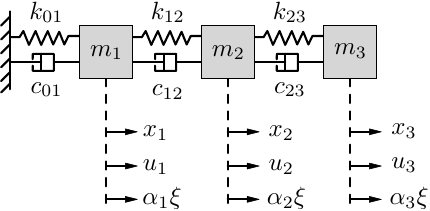}
	};
	\node[inner sep=0pt,anchor=south west] at (6.1cm,0.6cm){
		\includegraphics[width=3.7cm]
		{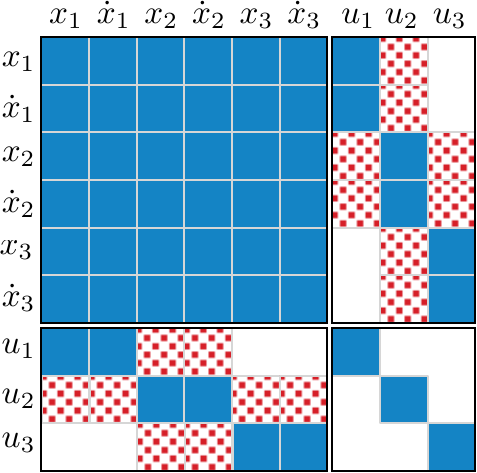}
	};
	%
	\node[inner sep=0pt,anchor=south west] at (10.9cm,0.6cm){
		\includegraphics[width=6.8cm]
		{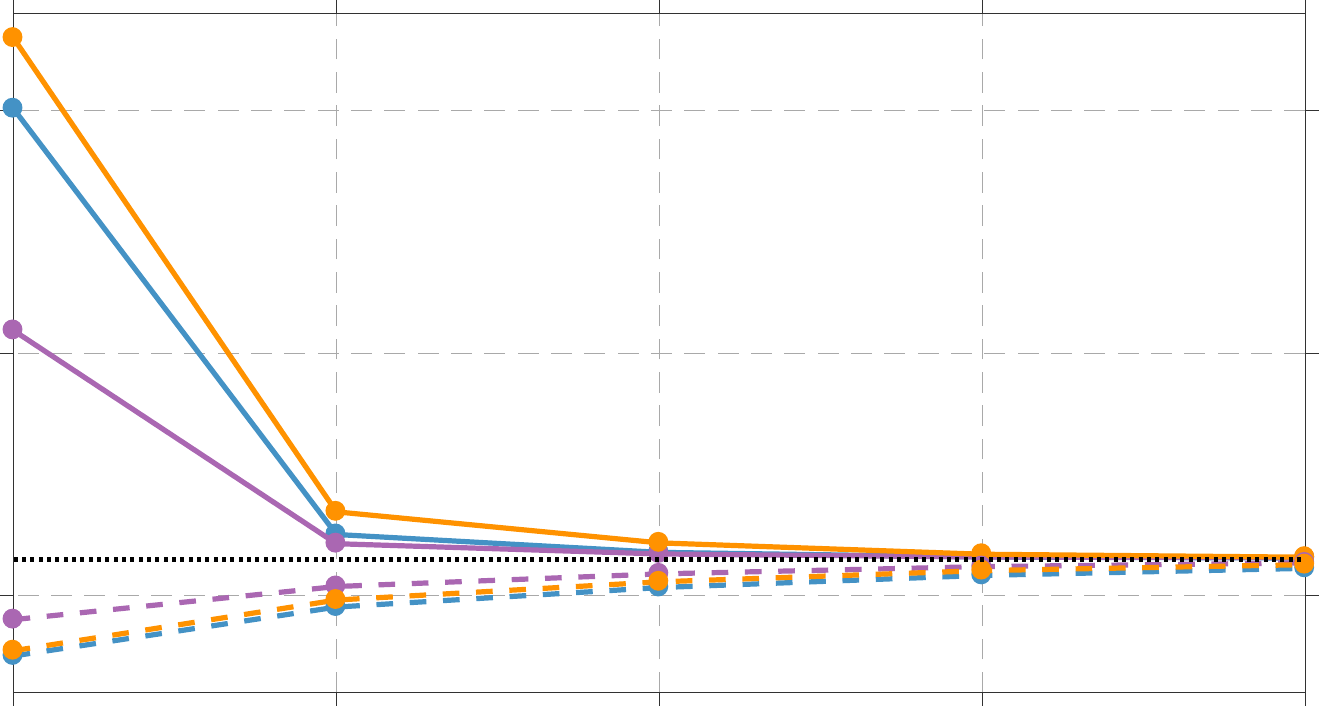}
	};
	%
	%
	\node[align=center , anchor=center , rotate=0] at (2.8cm,0.0cm) {\small{ (a) }};
	\node[align=center , anchor=center , rotate=0] at (7.95cm,0.0cm) {\small{ (b) }};
	\node[align=center , anchor=center , rotate=0] at (10.5cm,0.0cm) {\small{ (c) }};
	%
	\coordinate (GO) at (9.64cm,-0.2cm);
	\node[align=center , rotate=0] at ($(GO)+(4.68cm,0.2cm)$)
	{\small{ Number neighbour communication }};
	\node[align=center , anchor=north , rotate=0] at ($(GO)+(1.34cm,0.85cm)$) {\footnotesize{ $0$ }};
	\node[align=center , anchor=north , rotate=0] at ($(GO)+(3.01cm,0.85cm)$) {\footnotesize{ $1$ }};
	\node[align=center , anchor=north , rotate=0] at ($(GO)+(4.68cm,0.85cm)$) {\footnotesize{ $2$ }};
	\node[align=center , anchor=north , rotate=0] at ($(GO)+(6.34cm,0.85cm)$) {\footnotesize{ $3$ }};
	\node[align=center , anchor=north , rotate=0] at ($(GO)+(7.98cm,0.85cm)$) {\footnotesize{ $4$ }};
	\node[align=right , anchor=east , rotate=0] at ($(GO)+(1.34cm,1.40cm)$) {\footnotesize{ $750$ }};
	\node[align=right , anchor=east , rotate=0] at ($(GO)+(1.34cm,2.62cm)$) {\footnotesize{ $800$ }};
	\node[align=right , anchor=east , rotate=0] at ($(GO)+(1.34cm,3.86cm)$) {\footnotesize{ $850$ }};
	\draw[gray70,line width = 0.5pt,fill=white] ($(GO)+(3.3cm,2.9cm)$) rectangle ($(GO)+(5.7cm,4.3cm)$);
	\draw[matlabblue,line width = 1.2pt,solid]  ($(GO)+(3.5cm,4.0cm)$) -- ($(GO)+(3.9cm,4.0cm)$);
	\draw[matlabblue,line width = 1.2pt,dashed] ($(GO)+(4.0cm,4.0cm)$) -- ($(GO)+(4.4cm,4.0cm)$);
	\node[right] at ($(GO)+(4.4cm,4.0cm)$) {\color{matlabblue}{{\small$M\!=\!1$}}};
	\draw[matlabpurple,line width = 1.2pt,solid]  ($(GO)+(3.5cm,3.6cm)$) -- ($(GO)+(3.9cm,3.6cm)$);
	\draw[matlabpurple,line width = 1.2pt,dashed] ($(GO)+(4.0cm,3.6cm)$) -- ($(GO)+(4.4cm,3.6cm)$);
	\node[right] at ($(GO)+(4.4cm,3.6cm)$) {\color{matlabpurple}{\small$M\!=\!10$}};
	\draw[matlaborange ,line width = 1.2pt,solid]  ($(GO)+(3.5cm,3.2cm)$) -- ($(GO)+(3.9cm,3.2cm)$);
	\draw[matlaborange ,line width = 1.2pt,dashed] ($(GO)+(4.0cm,3.2cm)$) -- ($(GO)+(4.4cm,3.2cm)$);
	\node[right] at ($(GO)+(4.4cm,3.2cm)$) {\color{matlaborange}{\small$M\!=\!20$}};
	\end{tikzpicture}
	%
	%
	%
	\caption[Short-hand caption]{
		(a) Schematic of coupled oscillator model used to demonstrate using \textQ-functions for distributed control. The constituent sub-systems are the masses $m_i$, that respectively make decision $u_i$ based on state measurements $x_i$ and $\hat{x}_i$ in the decentralized setting, and may also have access to the state measurements of neighbouring masses in the distributed setting.
		%
		(b) Quadratic approximate \textQ-function structure used for the $P^Q$ matrix from \eqref{eq:quadratic_basis_function_space}, i.e., each square represents the coefficient of an order 2 monomial. When only the dark shaded elements are non-zero the greedy policy is decentralized, and when additionally the dotted elements are non-zero the greedy policy is distributed with nearest neighbour communication.
		%
		(c) Online performance (solid) and lower bounds (dashed) for the coupled oscillator example versus the communication connections. The centralized optimal (dotted) is 757.4 for this example. The horizontal axis is the number of neighbouring oscillators, in each direction, from which state measurements are available for making control decisions. Thus, $0$ represents a decentralized controller, and $1$ represents a controller with nearest neighbour communication.
	}
	\label{fig:smd}
\end{figure*}
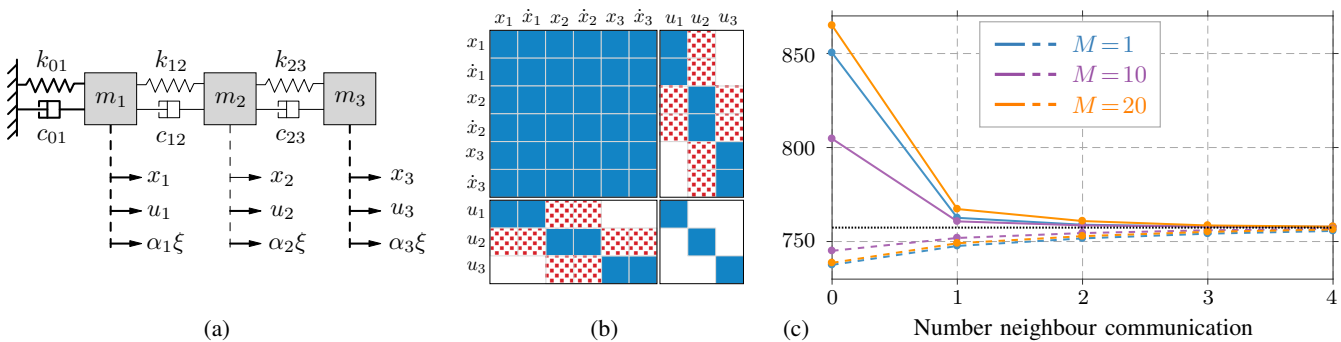

%
%

The most striking feature of this numerical example is the similarity between the iterated greedy policy and the MPC controller used for comparison. They only differ in the time horizon and choice of terminal cost function. The results in Table \ref{tab:numerical_results_NDexample} highlight that the $\hat{V}^\ast$ and $\hat{Q}^\ast$ encode a sufficient approximation of the cost-to-go function to allow for a shorter horizon to be used; for example, using $D=4$ for the iterated greedy policy results in comparable performance to an MPC controller with horizon $\smash{T\!=\!10}$ and an LQR based terminal cost, but at a fraction of the computational cost. In all cases, computing the policy involves solving a Quadratic Program with the number of decision variables and constraints proportional to the prediction horizon.

This numerical example also indicates that for a system where the input constraints are active at the end of the MPC prediction horizon, choosing an approximate value function or approximate \textQ-function for the terminal cost can lead to improved online performance. This comes at the expense of solving the approximate LP \eqref{eq:LP_approach_to_ADP_iterated_Qform} or \eqref{eq:LP_approach_to_ADP_iterated_Vform}, which for larger $M$ is more computationally demanding than solving the Riccati equation.
%
The computation time was 20 seconds for $\smash{M\!=\!1}$ and 24 minutes for $\smash{M\!=\!50}$ on a 4.0Ghz Intel Core i7 processor, with Appendix \ref{app:Sprocedure_reformulation} providing details of how the problem was reformulated for a commercial solver.
As the approximate LP only needs to be solved once for a particular system, this computation can be performed off-line, and the result offers improvements for the online performance and computation as demonstrated by this example.

%
%
%

\subsection{Coupled Oscillator Example} \label{sec:numerical_smd}

To demonstrate the application of \textQ-functions for distributed control we use a string of coupled oscillators, visualized as a spring-mass-dampener system in Fig.\,\ref{fig:smd}(a). Each mass is considered as a separate system, and needs to make its control decision based on the measurement of its own state, and possibly that of its nearest neighbours.

The coupled oscillator can be modelled by a linear system readily derived by writing the equations of motion for each mass. The state vector is the position and velocity of each mass, denoted as $x_i$ and $\dot{x}_i$ respectively. Each mass can be controlled by a driving force $\smash{u_i \!\in\! \rdim{}}$ applied to the mass. The exogenous driving force is $\smash{\xi \!\in\! \rdim{}}$ and the factor $\smash{\alpha_i \!\in\! \rdim{} }$ represents an external influence. The spring constant and dampening ratio of the elements connecting mass $i$ to mass $j$ are denoted by $k_{ij}$ and $c_{ij}$ respectively. The fixed wall is represented as $i=0$.

The online performance of the distributed control policies is compared to the optimal centralized policy. To be able to compute the centralized optimal, we use a quadratic stage and unconstrained state and action spaces. The stage cost for each mass is $l_i(x_i,\dot{x}_i,u_i) = 0.5 x_i^2 + \dot{x}_i^2 + 0.2 u_i^2$, with a discount factor of $\gamma = 0.99$, and the dynamics is converted to discrete time with a $0.05$ second sampling time. 
Fig.\,\ref{fig:smd}(a) shows a system with $3$ masses for clarity, but for the numerical results in Fig.\,\ref{fig:smd}(c) we simulate a system with $20$ masses.

As described in Section \ref{sec:unify_structured_functions}, an approximate \textQ-function can lead to a decentralized greedy policy if given an appropriate structure. The $Q^\ast$ is quadratic for this coupled oscillator example and due to the dynamic coupling the optimal greedy policy does not have a separable structure. Thus, for the restricted function space $\approxFuncSpaceXU$ we use quadratic functions parameterized as in \eqref{eq:quadratic_basis_function_space} with the $P^Q$ matrix restricted to have the structure shown in Fig.\,\ref{fig:smd}(b). The structure is shown for a three mass example and is readily extended for a longer string of masses. The approximate greedy policy is decentralized if the shaded structure is used, and distributed with nearest neighbour communication if additionally the dotted elements are non-zero. Note that $Q^\ast \notin \approxFuncSpaceXU$.

Fig.\,\ref{fig:smd}(c) presents the online performance results of using structured approximate \textQ-functions for decentralized and distributed control of the coupled oscillator system with 20 masses and the parameters randomly drawn from a uniform distribution on the following ranges: $\smash{ m_{i} \!\in\! \left[ 0.5  , 1.5  \right]}$, $\smash{k_{ij} \!\in\! \left[ 3.0 , 4.0 \right]}$, $\smash{c_{ij} \!\in\! \left[ 0.01 , 0.05 \right]}$, $\smash{\alpha_i \!\in\! \left[ 0.04 , 0.08 \right]}$.
%
The exogenous disturbance and initial condition are assumed to be distributed according to $\smash{\xi_t \!\sim\! \mcal{N}(0,1)}$, $\smash{x_i \!\sim\! \mcal{N}(0,0.5)}$, and $\smash{\dot{x}_i \!\sim\! \mcal{N}(0,1)}$ respectively.

We solve \eqref{eq:PropOfEquiv_forV} with $M\!=\!\left\{ 1 ,10 , 20 \right\}$ and simulate greedy policy \eqref{eq:approxpolicy_Qform}. The online performance is computed using $5000$ Monte Carlo samples from the initial state distribution, and the expectation with respect to $\xi$ is computed from $500$ Monte Carlo simulations each of length $2000$ time steps.
As a datum, the online performance of the centralized LQR controller is 757.4 which lies between the upper and lower bound curves in Fig.\,\ref{fig:smd}(c).
For each approximate \textQ-function, the lower-bound is computed from the same initial condition samples. Note that these are all lower bounds on the centralized LQR performance because problem \eqref{eq:PropOfEquiv_forV} is formulated to approximate the centralized problem.

The results in Fig.\,\ref{fig:smd}(c) show that, for this example, the decentralized/distributed ADP approach using \textQ-functions, can produce near centralized optimal performance: within $6.3\%$ in the decentralized case, within $1.4\%$ in the distributed, nearest neighbour communication, case. The online performance is significantly influenced by the choice of $M$.

%% file: sec/10_conclusion.tex

\section{Conclusions} \label{sec:conclusion}

In this paper we derived theoretical performance guarantees for the Linear Programming Approach to Approximate Dynamic Programming in continuous spaces.
We analyzed an iterated version of the greedy policy to provide a guarantee that the online performance of the policy is bounded.
We provided a Lyapunov-based bound on the approximation quality of a solution using the LP approach with the iterated Bellman inequality. This bound demonstrates a $61\%$ tightening, compared to the bound presented in \cite{boyd_iteratedBellman}, on the numerical example for which the bounds were evaluated.

We proposed a condition that allows for a more efficient iterated \textQ-function formulation. A numerical case study on linear-quadratic examples with a 50-dimensional state vector demonstrates the potential for large-scale systems.
Using an approximate value function or \textQ-function as a terminal cost for an MPC type controller achieves for these examples comparable online performance with one fifth of the online computational load.
The proposed condition applies also when using \textQ-functions in a decentralized control framework. The online performance using decentralized \textQ-functions is within $6.3\%$ of the optimal centralized performance for the coupled oscillator example.

As future work, we aim to adapt the LP approach to ADP to address the challenge of tuning the relevance weighting parameter, and through this reduce the conservativeness of the theoretical guarantees.
%
The numerical results demonstrate potential benefits for application to physical systems, particularly for systems with fast dynamics where the online policy is implemented on an embedded micro-controller with limited computation resources.

%% file: sec/11_appendix_A_LPequiv_Qform.tex


\section{Equivalence of \textQ-function LP} \label{app:LPequiv_Qform}

The theorem below provides the condition on $c(\cdot,\cdot)$ under which the solution of  \eqref{eq:bellman_Qform}, the \textQ-function variant of the Bellman equation, is feasible and optimal for the LP \eqref{eq:LP_approach_to_DP_Qform}.
Let $K(\cdot|\cdot,\cdot)$ denote the discrete-time transition kernel describing the state evolution under the dynamics and the exogenous and control inputs, i.e., given a Borel set $B \in \mcal{B}(\mcal{X})$,
	\begin{equation} \nonumber
		\begin{aligned}
			K\left(\, B \,\mid\,  x_t \,, u_t \,\right)
				\,=&\, \mrm{P}\left[\, g(x_t,u_t,\xi_t) \in B \,\right]
		\end{aligned}
	\end{equation}
represents the probability that state $x_{t+1}$ will be in $B$ given that the system is currently in state $x_t$ and input $u_t$ is played.
%
Furthermore, let $\smash{\mcal{M}(\spaceXbyU)}$ and $\smash{\mcal{M}(\spaceX)}$ be defined as the vector spaces of finite signed measures on $\smash{\spaceXbyU}$ and $\spaceX$ respectively, bounded as per \cite[Definition 6.3.2, 6.3.4]{hernandez_2012_discreteTimeMCP}.
%
We use $\mu(\intd{\cdot})$ throughout as short-hand notation for $\mu(\cdot) \intd{\cdot}$.

Given $\rho\in\mcal{M}(\spaceXbyU)$ we define an operator $T$ as,
	\begin{equation} \nonumber 
		\begin{aligned}
			\left( T \rho \right)(B) = \rho(B) - \gamma \, \int\nolimits_{\spaceXbyU} \, K\left( B \middle| x, u \right) \, \rho(\intd{(x,u)})
			\,.
		\end{aligned}
	\end{equation}
Thus, the dual LP of \eqref{eq:LP_approach_to_DP_Qform} is,
\begin{equation} \label{eq:LP_approach_to_DP_Qform_dual}
	\begin{aligned}
		\min\limits_{\mu} \quad&
			\int\nolimits_{\mcal{X} \times \mcal{U}} l(x,u) \, \mu(\intd{(x,u)})
			+ \int\nolimits_{\mcal{X} \times \mcal{U}} l(x,u) \, c(\intd{(x,u)})
		\\
		\text{s.t.} \quad& \left( T \mu \right) (B) = \gamma \int\limits_{\mcal{X} \times \mcal{U}} K(B|x,u) c(\intd{(x,u)})
			,\; \forall B \!\in\! \mcal{B}(\mcal{X})
		\\
		& \mu \in \mcal{M}^+(\spaceXbyU) ,\; (\mu+c) \in \mcal{M}^+(\spaceXbyU)
			\,.
	\end{aligned}
\end{equation}
where $\mcal{M}^+(\spaceXbyU)$ is the non-negative variant of $\mcal{M}(\spaceXbyU)$.
Now we state the requirement on the state-by-input relevance weighting for  \eqref{eq:LP_approach_to_DP_Qform} to recover $Q^\ast$ for $c$-a.a. $\xinX$.

\vspace{0.1cm}

\begin{theorem} \label{theorem:Q_primal_equiv}
	Under \cite[Assumptions 4.2.1(a), 4.2.1(b), 4.2.2]{hernandez_2012_discreteTimeMCP}, if $\nu$ and $c$ satisfy,
		\begin{equation} \label{theorem:Q_primal_equiv_eq1}
			\begin{aligned}
				\gamma \, \int\nolimits_{\mcal{X} \times \mcal{U}} \, K(B|x,u) \, c(\intd{(x,u)}) = \nu(B)
					,\; \forall\, B \in \mcal{B}(\mcal{X})
			\end{aligned}
		\end{equation}
	with $c \in \mcal{M}^+(\spaceXbyU)$, then,
	the optimal values of \eqref{eq:LP_approach_to_DP_Qform} and \eqref{eq:LP_approach_to_DP_Qform_dual} coincide with,
	\begin{equation} \nonumber
		\begin{aligned}
			\int\nolimits_{\spaceXbyU} Q^\ast(x,u) \, c(x,u) \, \intd{x} \intd{u}
			\,.
		\end{aligned}
	\end{equation}
\end{theorem}

\vspace{0.1cm}

\begin{IEEEproof}
	As the term $\int l \intd{c}$ in the objective of \eqref{eq:LP_approach_to_DP_Qform_dual} is an additive constant with respect to the decision variable $\mu$, we have by \cite[Theorem 6.3.7]{hernandez_2012_discreteTimeMCP} that the optimal value of \eqref{eq:LP_approach_to_DP_Qform_dual}, denoted $J_{\eqref{eq:LP_approach_to_DP_Qform_dual}}$, satisfies,
		\begin{equation} \nonumber
			\begin{aligned}
				&\, J_{\eqref{eq:LP_approach_to_DP_Qform_dual}}
				\\
				=&\, \int\nolimits_{\mcal{X}} V^\ast(x) \nu(\intd{x})
					+ \int\nolimits_{\mcal{X} \times \mcal{U}} l(x,u) \, c(\intd{(x,u)})
				\\
				=&\, \int\nolimits_{\mcal{X} \times \mcal{U}} \, \left( \gamma \int\nolimits_{y\in\mcal{X}}  V^\ast(y) K(\intd{y}|x,u) + l(x,u) \right) \, c(\intd{(x,u)})
				\\
				=&\, \int\nolimits_{\mcal{X} \times \mcal{U}} \, \Big(\, l(x,u) \,+\, \gamma \, \expval{\xi}{ V^\ast(g(x,u,\xi))} \,\Big) \, c(\intd{(x,u)})
				\\
				=&\, \int\nolimits_{\mcal{X} \times \mcal{U}} \, \Big(\, Q^\ast(x,u) \,\Big) \, c(\intd{(x,u)})
					\,.
			\end{aligned}
		\end{equation}
	The first equality follows from \cite[Theorem 6.3.7]{hernandez_2012_discreteTimeMCP}. The second equality substitutes \eqref{theorem:Q_primal_equiv_eq1} for $\nu$ and uses Fubini's theorem to switch the order of integration. The third equality is the definition of integration with respect to the transition kernel, and the final equality is the definition of $Q^\ast$.
	
	The strong duality between \eqref{eq:LP_approach_to_DP_Qform_dual} and \eqref{eq:LP_approach_to_DP_Qform} follows from \cite[Theorem 6.3.8]{hernandez_2012_discreteTimeMCP}. They use the sequence of value functions: $v_0(\cdot)=0$ and for all $n \geq 1$ and $x\in\mcal{X}$
		\begin{equation} \nonumber
			\begin{aligned}
				v_{n}(x) = \min_{u\in\mcal{U}} \left[ l(x,u) + \gamma  \int\nolimits_{y\in\mcal{X}} v_{n-1}(y) \, K\left( \intd{y} \middle| x,\pi(x) \right) \right]
				\,,
			\end{aligned}
		\end{equation}
	see \cite[equation 6.3.38]{hernandez_2012_discreteTimeMCP}, to show that both the primal and dual programs converger to $V^\ast$.
	By defining a \textQ-function to match each value function for all $n \geq 0$ and $(x,u) \!\in\! \spaceXbyU$
		\begin{equation} \nonumber
			\begin{aligned}
				q_{n}(x,u) := l(x,u) + \gamma \int\nolimits_{y\in\mcal{X}} \, v_{n}(y) \, K(\intd{y}|x,u)
					\,,
			\end{aligned}
		\end{equation}
	it follows that problems \eqref{eq:LP_approach_to_DP_Qform} and \eqref{eq:LP_approach_to_DP_Qform_dual} have the same optimal value, and that problem \eqref{eq:LP_approach_to_DP_Qform} attains the supremum.
\end{IEEEproof}

%% file: sec/11_appendix_B_LPreformulation_propositions.tex

\section{Propositions for LP reformulation} \label{app:LPreformulation_propositions}

This appendix states the propositions necessary for the reformulation of the non-linear iterated $F$-operator inequality constraint as a set of linear constraints.

\vspace{0.1cm}
\begin{proposition} \label{proposition:F_operator_inequality_reformulation}
	For an arbitrary $Q:\spaceXbyU \!\rightarrow\! \rdim{}$ the following statements are equivalent:
	\vspace{0.1cm}
	\begin{enumerate}
		\renewcommand{\labelenumi}{(\roman{enumi})}
		\item $Q(x,u) \leq F Q(x,u)$ for all $\xinX$ and all $\uinU$;
		
		\vspace{0.1cm}
		
		\item There exists $V \!\in\! \mcal{F}(\mcal{X})$ such that $Q(x,u) \leq \mcal{T}_u V(x)$ and  $V(x) \leq Q(x,u)$ for all $\xinX$ and all $\uinU$.
	\end{enumerate}
\end{proposition}
\vspace{0.1cm}
The proof is given in \cite[Theorem 2]{vanroy_decentADP}.
%
Note, if $V$ is in some subset of $\smash{\funcSpaceX}$, then the reformulation is only sufficient, i.e., (ii)$\,\Rightarrow\,$(i).

\vspace{0.1cm}
\begin{proposition} \label{proposition:F_operator_iterated_inequality_reformulation}
	For an arbitrary $Q:\spaceXbyU \rightarrow \rdim{}$ the following are equivalent:
	\vspace{0.1cm}
	\begin{enumerate}
		\renewcommand{\labelenumi}{(\roman{enumi})}
		\item $Q(x,u) \leq F^M Q(x,u)$ for all $\xinX$ and all $\uinU$;
		
		\vspace{0.1cm}
		
		\item There exists $Q_1 , \dots , Q_{\smash{M-1}} \!\in\! \funcSpaceXU$ such that:
		\begin{equation} \nonumber
			\begin{aligned}
				Q(x,u) &\leq F Q_{1}(x,u)
				\,,
				\\
				Q_{j-1}(x,u) &\leq F Q_{j}(x,u) 
				\,,\quad j \!=\! 2,\dots,M\!-\!1
				\,,
				\\
				Q_{M-1}(x,u) &\leq F Q(x,u)
				\,,
			\end{aligned}
		\end{equation}
		
	\end{enumerate}
	where the inequalities hold for all $\xinX$ and all $\uinU$.
\end{proposition}
\vspace{0.1cm}
The proof follows from \cite[\S 3.4]{boyd_iteratedBellman}.
%
Note, if for any $j$, $Q_j$ is in some subset of $\funcSpaceXU$, then the reformulation is only sufficient, i.e., (ii)$\,\Rightarrow\,$(i).

%% file: sec/11_appendix_C_bound_online_perf_kernel_intro.tex


\section{Proof of online performance bound} \label{app:proof_bound_online_performance}

Given a measure $\rho\in\mcal{M}(\mcal{X})$ (see \cite[Definition 6.3.4]{hernandez_2012_discreteTimeMCP}), a feasible policy $\pi:\mcal{X} \rightarrow \mcal{U}$, and a Borel set $B \in \mcal{B}(\mcal{X})$, define the operator $T_{\pi}: \mcal{M}(\mcal{X}) \rightarrow \mcal{M}(\mcal{X})$ as,
	\begin{equation} \nonumber
		\begin{aligned}
			\left(\, T_{\pi} \, \rho \,\right)(B) \,=\, \rho(B) \,-\, \gamma \, \int\nolimits_{x\in\mcal{X}} \, K\left( B \,\middle|\, x,\pi(x) \right) \, \rho(\intd{x})
			\,.
		\end{aligned}
	\end{equation}
Thus $T_{\pi}$ represents the discounted difference in occupancy measure between two time steps of the stochastic process. Given a function $V\in\mcal{F}(\mcal{X})$, and the same feasible policy, consider also the operator $T_{\pi}^{\ast}: \mcal{F}(\mcal{X}) \rightarrow \mcal{F}(\mcal{X})$ defined as,
	\begin{equation} \label{eq:lasserre_T_pi_operator_for_functions}
		\begin{aligned}
			\left( T_{\pi}^{\ast}  V \right)\!(x)
				= V(x) - \gamma  \int\nolimits_{y\in\mcal{X}} V(y) \, K\left( \intd{y} \middle| x,\pi(x) \right)
				.
		\end{aligned}
	\end{equation}
Thus $T_{\pi}^\ast$ represents the expected value of discounted difference between two time steps of the stochastic process. Both operators define a continuous linear map on the corresponding spaces and are adjoints of each other, i.e.,
	\begin{equation} \label{eq:appendix_operators_adjoint} 
		\begin{aligned}
			\int\nolimits_{x\in\mcal{X}} V(x) \left( T_{\pi}  \rho \right)(\intd{x})
				=
				\int\nolimits_{x\in\mcal{X}} \left( T_{\pi}^{\ast}  V \right)(x) \, \rho(\intd{x})
				\,,
		\end{aligned}
	\end{equation}
see \cite[Section 6.3]{hernandez_2012_discreteTimeMCP}. The online performance bound for finite space is proven by inverting the transition kernel matrix, see \cite[Theorem 1]{vanRoy_linApproxDP}. The adjoint property of $T_{\pi}$ and $T_{\pi}^\ast$ can be seen as a counterpart to inverting the transition kernel.

A required identity is that the online performance can be expressed in terms of the stage cost and the frequency measure defined in Section \ref{sec:bounds_online}. Given a policy, $\pi:\mcal{X} \rightarrow \mcal{U}$ and the expected state frequency with respect to that policy, $\tilde{\mu}$, the online performance is expressed as:
	\begin{equation} \label{eq:appendix_online_perf_measure_identity} 
		\begin{aligned}
			V_{\pi}(y) \,:=&\, \mbb{E}\left[\, \sum\nolimits_{t=0}^{\infty} \gamma^t l(x_t,\pi(x_t))  \,\middle|\, x_0 = y\,\right]
			\\
			=&\, \int_{\mcal{X}} \, l(x,\pi(x)) \, \tilde{\mu} \left(\intd{x}\right)
		\end{aligned}
	\end{equation}
When the left hand side is integrated over the initial state distribution, $\nu$, then $\tilde{\mu}$ is chosen accordingly.

A final identity relates the initial state distribution to the expected state frequency. Given any $B \in \mcal{B}(\mcal{X})$ the following relation holds:
	\begin{equation} \label{eq:appendix_measures_identity} 
		\begin{aligned}
			\nu(B) =&\, \tilde{\mu}(B) \,-\, \gamma \, \int_{x\in\mcal{X}} \, K\left(\, B \,\middle|\, x,\pi(x) \,\right) \, \tilde{\mu}\left(\intd{x}\right)
			\\
			=&\, \left(\, T_{\pi} \, \tilde{\mu} \,\right)(B)
		\end{aligned}
	\end{equation}
 This identity stems from \cite[eq. (6.3.10)]{hernandez_2012_discreteTimeMCP}.

\vspace{0.2cm}

We now have all the tools required to prove Theorem \ref{theorem:online_performance_bound_iterated_Qform}.

%% file: sec/11_appendix_C_bound_online_perf_proof_Qform.tex


\vspace{0.2cm}
\begin{IEEEproof}[Proof of Theorem \ref{theorem:online_performance_bound_iterated_Qform}]
		
	For all $k\in\mbb{N}$,
	\begin{equation} \label{eq:online_performance_bound_iterated_Qform_01}
		\begin{aligned}
			\hat{Q}(x,u) \leq F^k \hat{Q}(x,u) \leq Q^\ast(x,u) \leq Q_{\hat{\pi}}(x,u)
				\,,
		\end{aligned}
	\end{equation}
	for all $\xinX$ and $\uinU$, and hence also for all $u\!=\!\hat{\pi}(x)\!\in\!\mcal{U}$.

	Recalling the notation $Q|_{\pi}(x) := \smash{Q(x,\pi(x))}$, we have,
	\begin{equation} \nonumber
		\begin{aligned}
			\hspace{-0.3cm}
			&\, \left\| \, V_{\hat{\pi}} \,-\, V^\ast \, \right\|_{1,\nu}
			\\
			=&\, \int_{\mcal{X}} \, \left( V_{\hat{\pi}}(x) \,-\, V^\ast(x) \, \right) \nu(\intd{x})
			\\
			\leq&\, \int_{\mcal{X}} \, \left( V_{\hat{\pi}}(x) \,-\, \kcol{\left.\left( F^{D} \hat{Q}\right)\right|_{\hat{\pi}}}(x) \, \right) \nu(\intd{x})
			\\
			=&\, \kcol{\int\limits_{\mcal{X}} \, l(x,\hat{\pi}(x)) \, \tilde{\mu}(\intd{x})}
				- \int\limits_{\mcal{X}} \, 
					\left.\left( F^{D} \hat{Q}\right)\right|_{\hat{\pi}}(x)
					\,
					\kcol{ \left( T_{\pi} \tilde{\mu} \right) (\intd{x}) }
			\\
			=&\, \int\limits_{\mcal{X}} \, l(x,\hat{\pi}(x)) \, \tilde{\mu}(\intd{x})
				- \int\limits_{\mcal{X}} \,
					\left( \kcol{ T_{\pi}^{\ast} } \, \left.\left( F^{D} \hat{Q}\right)\right|_{\hat{\pi}} \right)(x)
					\,
					\kcol{\tilde{\mu}(\intd{x})}
			\\
			=&\, \int\limits_{\mcal{X}} \kcol{ \left.\left( F^{D+1} \hat{Q}\right)\right|_{\hat{\pi}}(x) } \, \tilde{\mu}(\intd{x})
				\,-\, \int\limits_{\mcal{X}} \left.\left( F^{D} \hat{Q}\right)\right|_{\hat{\pi}}(x) \, \tilde{\mu}(\intd{x})
			\\
			\kcol{\leq}&\, \int\limits_{\mcal{X}} \kcol{ \left.Q^{\ast}\right|_{\hat{\pi}}(x) } \, \tilde{\mu}(\intd{x})
				\,-\, \int\limits_{\mcal{X}} \left.\left( F^{D} \hat{Q}\right)\right|_{\hat{\pi}}(x) \, \tilde{\mu}(\intd{x})
			\\
			=&\, \frac{1}{1-\gamma} \,\, \left\|\, \left.Q^{\ast}\right|_{\hat{\pi}} \,-\, \left.\left( F^{D} \hat{Q}\right)\right|_{\hat{\pi}} \,\right\|_{1,(1-\gamma)\tilde{\mu}}
		\end{aligned}
	\end{equation}
	
	The first equality and first inequality hold by the point-wise ordering of \eqref{eq:online_performance_bound_iterated_Qform_01}. The second equality uses \eqref{eq:appendix_online_perf_measure_identity} for the first term and \eqref{eq:appendix_measures_identity} for the second term. The third equality uses \eqref{eq:appendix_operators_adjoint}, while the fourth uses \eqref{eq:lasserre_T_pi_operator_for_functions} to expand the $T_\pi^\ast$ operator, and then the definition of the $F$-operator and the chosen policy to construct the first term. The last inequality and equality follow from the point-wise ordering of \eqref{eq:online_performance_bound_iterated_Qform_01} and the definition of the 1-norm. The factor $(1\!-\!\gamma)$ was introduced so that the scaling in the 1-norm is a probability measure.
\end{IEEEproof}

\vspace{0.2cm}

%% file: sec/11_appendix_D_bound_inf_norm.tex


\section{Proof of Infinity-norm bound} \label{app:proof_bound_inf_norm_Qform}

The proof of Theorem \ref{theorem:inf_norm_bound_iterated_Qform} uses two additional lemmas that are presented first, and then we present the proof of Theorem \ref{theorem:inf_norm_bound_iterated_Qform}.
%
Lemma \ref{Qform_constraint_violation_infinity_norm} provides a point-wise bound on how much the $M$-iterated $F$-operator inequality is violated for any given $\mcal{Q}$ function, from the restricted function space or otherwise. This is used in the proof of Lemma \ref{lemma:Q_feasible_after_downshift}, which shows that given a $\smash{\hat{Q} \in \hat{\mcal{F}}(\mcal{X} \!\times\! \mcal{U})}$, it can be downshifted by a certain constant amount to satisfy the iterated $F$-operator inequality. The constant by which it is downshifted relates directly to the constant on the right-hand-side of Theorem \ref{theorem:inf_norm_bound_iterated_Qform}.
%
The proof here is an adaptation to $\mathcal{Q}$-functions of the proof for Value functions that is given in \cite[\S 4.3]{boyd_iteratedBellman}.

\vspace{0.1cm}

\begin{lemma} \label{Qform_constraint_violation_infinity_norm}
	For any $Q : \spaceXbyU \rightarrow \rdim{}$ and $M\in\mbb{N}$ iterations,
	\begin{equation} \nonumber
	\begin{aligned}
	\left( F^M Q \right)(x,u) \,\geq\, Q(x,u) \,-\, \left(1+\gamma^M\right) \, \left\| Q^\ast - Q \right\|_\infty
	\,,
	\end{aligned}
	\end{equation}
	for all $\xinX$ and all $\uinU$.
\end{lemma}

\vspace{0.1cm}

\begin{IEEEproof}
	Starting from the terms not involving $\gamma$,
	\begin{equation} \nonumber 
		\begin{aligned}
			&Q(x,u) \,-\, \left\|\, Q^\ast \,-\, Q \right\|_\infty \,-\, \left(\, F^M \, Q \,\right)(x,u)
			\\
			\leq&\, \kcol{Q^\ast(x,u)} \,-\, \left(\, F^M \, Q \,\right)(x,u)
				\,,\quad \forall \, x\in\mcal{X},\,u\in\mcal{U}
			\\
			\kcol{\leq}&\, \kcol{\left\|\, \kcol{\left(\, \kcol{F^M} \, Q^\ast \,\right) \,-\, \left(\, F^M \, Q \,\right)} \,\right\|_\infty}
			\\
			\kcol{\leq}&\, \kcol{\gamma^M} \, \left\|\, Q^\ast \,-\, Q \,\right\|_\infty
				\,.
		\end{aligned}
	\end{equation}
	The first inequality follows from the definition of the $\infty$-norm, and the second inequality comes from $Q^\ast(x,u)=(FQ^\ast)(x,u)$ and the $\infty$-norm definition. Finally, the third inequality is due to the $\gamma$-contractive property of the $F$-operator. Re-arranging, the result follows.
\end{IEEEproof}

\vspace{0.1cm}


\begin{lemma} \label{lemma:Q_feasible_after_downshift}
	Let $\hat{Q}(x,u) \in \approxFuncSpaceXU$ be an arbitrary element from the basis functions set, and let $\tilde{Q}(x,u)$ be defined as,
	\begin{equation} \label{lemma:Q_feasible_after_downshift_eq1}
		\begin{aligned}
			\tilde{Q}(x,u) = \hat{Q}(x,u) \,-\, \underbrace{ \frac{1+\gamma^M}{1-\gamma^M} \, \| Q^\ast - \hat{Q} \|_\infty}_{\text{downwards shift term}}
				\,,
		\end{aligned}
	\end{equation}
	then $\tilde{Q}(x,u)$ satisfies the iterated $F$-operator inequality, and if $\approxFuncSpaceXU$ allows for affine combinations of the basis functions, then $\tilde{Q}$ is also an element of $\approxFuncSpaceXU$.
\end{lemma}

\vspace{0.2cm}

\begin{IEEEproof}
	Let $\beta \in \rdim{}$ denote the constant \emph{downwards shift term} for notational convenience. Using the definition of the $F$-operator we see that for any function $Q(x,u)$,
	\begin{equation} \nonumber
		\begin{aligned}
			&\, \left(\, F \, \left(Q + \beta\right) \,\right)(x,u)
			\\
			=&\, l(x,u) \,+\, \gamma \, \min_{v\in\mcal{U}} \expval{}{ \, Q(f(x,u,\xi),v) \,+\, \beta \, }
			\\
			=&\, \left(F Q\right)(x,u) \,+\, \gamma \, \beta
				\,.
		\end{aligned}
	\end{equation}
	where the equalities hold for all $x\in\mcal{X}$, $u\in\mcal{U}$. The first equality comes from the definition of the $F$-operator, and the second equality holds as $\beta$ is an additive constant in the objective of the minimization.
	
	Iterating the same argumentation $M$-times leads to
	\begin{equation} \label{lemma:Q_feasible_after_downshift_eq2}
		\begin{aligned}
			&\, \left( F^M \left(Q + \beta \right) \right)(x,u)
			\\
			=&\, \left( F^{M-1} \left( F \left( Q + \beta \right) \right) \right)(x,u) 
			\\
			=&\, \left( F^{M-1} \left( \left(F Q \right) + \gamma \beta \right) \right)(x,u) 
			\\
			=&\, \left( F^{M-2} \left( \left(F^2 Q \right) + \gamma^2 \beta \right) \right)(x,u) 
			\\
			=&\, \dots
			\\
			=&\, \left(F^M Q\right)(x,u) + \gamma^M \, \beta
				\,,
		\end{aligned}
	\end{equation}
	where the equivalences hold point-wise for all $x \in \mcal{X}$, $u \in \mcal{U}$. Now we show that $\tilde{Q}$ satisfies the iterated $F$-operator inequality,
	\begin{equation} \nonumber
		\begin{aligned}
			&\, \left( F^M \tilde{Q} \right)(x,u)
			\\
			=&\, \left( F^M \hat{Q} \right)(x,u) \,-\, \gamma^M \left( \frac{1+\gamma^M}{1-\gamma^M} \, \| Q^\ast - \hat{Q} \|_\infty \right)
			\\
			\kcol{\geq}&\, \kcol{\hat{Q}(x,u) - \left(1+\gamma^M\right) \, \left\| Q^\ast - \hat{Q} \right\|_\infty}
				\\
				&\qquad\qquad-\, \gamma^M \left( \frac{1+\gamma^M}{1-\gamma^M} \, \| Q^\ast - \hat{Q} \|_\infty \right)
			\\
			=&\, \tilde{Q}(x,u)
				\,,
		\end{aligned}
	\end{equation}
	where the first equality comes from \eqref{lemma:Q_feasible_after_downshift_eq2}, the inequality is a direct application of Lemma \ref{Qform_constraint_violation_infinity_norm} to the term $(F^M \hat{Q})$ and holds for all $x\in\mcal{X}$, $u\in\mcal{U}$, and the final equality follows from \eqref{lemma:Q_feasible_after_downshift_eq1}.

	Finally, if $\hat{\mcal{F}}(\mcal{X} \times \mcal{U})$ allows for affine combinations of the basis functions, then $\hat{Q}\in\hat{\mcal{F}}(\mcal{X} \times \mcal{U})$ implies $\tilde{Q} \in \hat{\mcal{F}}(\mcal{X} \times \mcal{U})$ as the \emph{downward shift term} is an additive constant.
\end{IEEEproof}

\vspace{0.2cm}


\begin{IEEEproof}[Proof of of Theorem \ref{theorem:inf_norm_bound_iterated_Qform}]
	
	Given any $\smash{\hat{Q}\in\approxFuncSpaceXU}$, construct $\smash{\tilde{Q} \in \approxFuncSpaceXU}$ following Lemma \ref{lemma:Q_feasible_after_downshift} to be feasible for the approximate iterated LP.
	Working from the left hand side of equation \eqref{eq:theorem:inf_norm_bound_iterated_Qform_bound},
	\begin{equation} \nonumber
		\begin{aligned}
			&\, \left\|\, Q^\ast \,-\, \hat{Q}^\ast \,\right\|_{1,c(x,u)}
			\\
			\kcol{\leq}&\, \left\|\, Q^\ast \,-\, \kcol{\tilde{Q}} \,\right\|_{1,c(x,u)}
			\\
			\kcol{\leq}&\, \left\|\, Q^\ast \,-\, \tilde{Q} \,\right\|_{\kcol{\infty}}
			\\
			\kcol{\leq}&\, \left\|\, Q^\ast \,-\, \kcol{\hat{Q}} \,\right\|_\infty \,\kcol{+}\, \left\|\, \kcol{\hat{Q}} \,-\, \tilde{Q} \,\right\|_\infty
			\\
			=&\, \left\|\, Q^\ast \,-\, \hat{Q} \,\right\|_\infty \,+\, \kcol{\frac{1+\gamma^M}{1-\gamma^M} \, \left\|\, Q^\ast \,-\, \hat{Q} \,\right\|_\infty}
			\\
			=&\, \frac{2}{1\,-\,\gamma^M} \,\, \left\|\, Q^\ast \,-\, \hat{Q} \,\right\|_\infty
		\end{aligned}
	\end{equation}
	where the first inequality holds by Lemma \ref{lemma:approxLP_for_Q_solves_min_1norm} because $\tilde{Q}$ is also feasible for \eqref{eq:LP_approach_to_ADP_iterated_Qform}, the second inequality by assuming without loss of generality that $c(x,u)$ is a probability measure, the third inequality is an application of the triangle inequality, the first equality stems directly from the definition of $\tilde{Q}$, and the final is an algebraic manipulation.
	As this argumentation holds for any $\hat{Q}\in\hat{\mcal{F}}(\mcal{X} \times \mcal{U})$,  the result follows.
\end{IEEEproof}


%% file: sec/11_appendix_E_bound_lyap_Vform.tex


\section{Proof of Lyapunov-based bound} \label{app:proof_bound_lyap_Vform}


The proof of Theorem \ref{theorem:bound_lyapunov_iterated_Vform} uses four lemmas that are derived first, and then we present the proof of Theorem \ref{theorem:bound_lyapunov_iterated_Vform}.
%
Lemma \ref{lemma:T_iterated_V_diff_leq_H_iterated_V_diff} bounds the difference after applying $M$ iterations of the Bellman operator to 2 different Value functions. The bound is given by $M$ iterations of the $H_V$ operator introduced in Section \ref{sec:bounds_fitting_lyap} and is used in Lemma \ref{lemma:T_iterated_V_geq_V_minusLyap} to give a bound on how much the $M$-iterated Bellman inequality is violated for any given Value function. This constraint violation bound is given in terms of a Lyapunov function and is used in Lemma \ref{lemma:feasible_forIterated_V_by_shifting} to prove that given any $\hat{V} \in \approxFuncSpaceX$, it can be downshifted by a scalar multiple of a Lyapunov function to satisfy the $M$-iterated Bellman inequality. The Lyapunov function appearing in the downshift relates directly to the Lyapunov function and relevance weighting on the right-hand-side of the Theorem \ref{theorem:bound_lyapunov_iterated_Vform} bound.
The proof of Theorem \ref{theorem:bound_lyapunov_iterated_Vform} is reminiscent of that for \cite[Theorem 3]{vanRoy_linApproxDP}, but requires an adapted analysis for consideration of the iterated Bellman inequality and continuous spaces.

\vspace{0.2cm}

\begin{lemma} \label{lemma:T_iterated_V_diff_leq_H_iterated_V_diff}
	For any two functions $V_1,V_2 : \mcal{X} \rightarrow \rdim{}$,
	\begin{equation} \nonumber
			\left| (\mcal{T}^M V_1)(x) - (\mcal{T}^M V_2)(x) \right|
				\,\leq\,
				\gamma^M \, \left( H_V^M ( \left| V_1 - V_2 \right| ) \right)(x)
			\,,
	\end{equation}
	for all $\xinX$, and any $\smash{M \in \mathbb{N}}$.
\end{lemma}

\vspace{0.2cm}

\begin{IEEEproof}
	The lemma will be proven by induction. For $M=1$, we first show that the inequality hold without $|\cdot|$. Letting $u_1^\ast$ denote the minimizer for $\mcal{T}V_1$ and $u_2^\ast$ for $\mcal{T}V_2$,
	\begin{equation} \label{eq:lemma:T_iterated_V_diff_leq_H_iterated_V_diff_eq01}
		\begin{aligned}
			&\, (\mcal{T} V_1)(x) - (\mcal{T} V_2)(x)
			\\
			=&\, \left( \mcal{T}_u V_1 \right)(x,\bcol{u_1^\ast})
				\,-\, \left( \mcal{T}_u V_2 \right)(x,\bcol{u_2^\ast})
			\\
			\kcol{\leq}&\, \left( \mcal{T}_u V_1 \right)(x,\bcol{u_2^\ast})
				\,-\, \left( \mcal{T}_u V_2 \right)(x,\bcol{u_2^\ast})
			\\
			\kcol{\leq}&\, \gamma \, \bcol{\max_{u\in\mcal{U}}} \left(\, (\mcal{T}_u V_1)(x,\bcol{u}) - (\mcal{T}_u V_2)(x,\bcol{u}) \,\right)
			\\
			\kcol{\leq}&\, \gamma \, \max_{u\in\mcal{U}} \bcol{\big|\, \kcol{ \expval{}{V_1(f(x,u,\xi))} - \expval{}{V_2(f(x,u,\xi))} } \,\big|}
				\,,
		\end{aligned}
	\end{equation}
	where the inequalities hold for all $x \in \mcal{X}$. The first equality is the definition of $\mcal{T}$ in terms of $\mcal{T}_u$, and the first inequality holds by definition of $u_1^\ast$ being the minimizer for $\mcal{T}V_1$. The second inequality holds as the same $u_2^\ast$ appears in both terms. The final inequality holds by definition of $\mcal{T}_u$ and $|\cdot|$.

	An entirely analogous argument establishes that $(\mcal{T} V_2)(x) - (\mcal{T} V_1)(x)$ is bounded above by the same final term in \eqref{eq:lemma:T_iterated_V_diff_leq_H_iterated_V_diff_eq01}. Hence the result for $M=1$ follows as,
	\begin{equation} \label{eq:lemma:T_iterated_V_diff_leq_H_iterated_V_diff_eq02}
		\begin{aligned}
			&\, \left| (\mcal{T} V_1)(x) - (\mcal{T} V_2)(x) \right|
			\\
			\kcol{\leq}&\, \gamma \, \max_{u\in\mcal{U}} \big|\, \expval{}{V_1(f(x,u,\xi))} - \expval{}{V_2(f(x,u,\xi))} \,\big|
			\\
			\kcol{\leq}&\, \gamma \, \max_{u\in\mcal{U}} \, \bcol{\expval{}{\, \kcol{ \left| \, V_1(f(x,u,\xi)) \,-\, V_2(f(x,u,\xi)) \, \right| } \,} }
			\\
			=&\, \gamma \, \left( H_V \, \left( \left| \, V_1 \,-\, V_2 \right) \, \right| \, \right)(x)
				\,,
		\end{aligned}
	\end{equation}
	where the inequalities hold for all $x \in \mcal{X}$. The first inequality follows from \eqref{eq:lemma:T_iterated_V_diff_leq_H_iterated_V_diff_eq01}. The second inequality uses \cite[Lemma 1.7.2]{christensen_2010_functions} to exchange the expectation and absolute value. The final equivalence is the definition of $H_V$ as per Section \ref{sec:bounds_fitting_lyap}.

	Assume the statement holds true for some $k\in\mbb{N}$, i.e.,
	\begin{equation} \nonumber
		\begin{aligned}
			\left| \left( \mcal{T}^k V_1 \right)(x) - \left( \mcal{T}^k V_2 \right)(x) \right| \,\leq\, \gamma^k \, \left( H_V^k  \left( \left| V_1 - V_2 \right| \right) \right)(x)
				\,,
		\end{aligned}
	\end{equation}
	and show it therefore holds true for $k+1$:
	\begin{equation} \nonumber
		\begin{aligned}
			&\,\left| (\mcal{T}^{k+1} V_1)(x) - (\mcal{T}^{k+1} V_2)(x) \right|
			\\
			=&\,\left|\, \left(\, \bcol{\mcal{T}^{k}} \, \left(\, \bcol{\mcal{T}} V_1 \,\right) \,\right)(x) \,-\, \left(\, \bcol{\mcal{T}^{k}} \, \left(\, \bcol{\mcal{T}} V_2 \,\right) \,\right)(x) \,\right|
			\\
			\kcol{\leq}&\, \bcol{ \gamma^k \, \left(\, H_V^k \, \left(\, \left|\,  \kcol{ \left(\mcal{T} V_1\right) \,-\, \left( \mcal{T} V_2 \right) } \,\right| \, \right) \, \right) } (x)
			\\
			\kcol{\leq}&\, \gamma^{k} \,  \left(\, H_V^{k} \, \left(\, \bcol{ \gamma \, H_V \left(\, \left| V_1 - V_2 \right| \,\right) } \,\right) \right)(x)\,
			\\
			=&\, \bcol{\gamma^{k+1}} \,  \left(\, \bcol{H_V^{k+1}} \, \left(\, \left| V_1 - V_2 \right| \,\right) \,\right)(x)
		\end{aligned}
	\end{equation}
	where the inequalities hold for all $x \in \mcal{X}$. The first equivalence splits $\mcal{T}^{k+1}$ so that the induction assumption can be used to establish the first inequality. The second inequality uses \eqref{eq:lemma:T_iterated_V_diff_leq_H_iterated_V_diff_eq02} and the monotonicity property of $H_V^k$. The final equivalence follows by algebra.
	
	By induction the claim holds for any integer $M \geq 1$.
\end{IEEEproof}

\vspace{0.2cm}

\begin{lemma} \label{lemma:T_iterated_V_geq_V_minusLyap}
	For any positive function $V^+:\mcal{X} \rightarrow \rdim{}_{++}$, any function $V:\mcal{X} \rightarrow \rdim{}$, and any integer $M \geq 1$,
	\begin{equation} \nonumber
			V(x) \,-\, (\mcal{T}^M V)(x)
				\,\leq\,
				\left( V^+(x) + \gamma^M (H_V^M V^+)(x) \right) \, \epsilon
	\end{equation}
	for all $x \in \mcal{X}$, where $\epsilon = \left\| V^\ast - V \right\|_{\infty,1/V^+}$.
\end{lemma}

\vspace{0.2cm}

\begin{IEEEproof}
	First we find a relation between $V^+$, $V$, and $V^\ast$ based on the weighted infinity norm.
	\begin{equation} \label{eq:01_for_T_iterated_V_geq_V_minusLyap} 
		\begin{aligned}
			&\, \epsilon \, V^+(x) \,=\,
				\left\| V^\ast - V \right\|_{\infty,1/V^+} \, V^+(x)
			\\
			\kcol{\geq}&\, \left|\, V^\ast(\bcol{x}) - V(\bcol{x}) \,\right| \, \left(\, 1 / V^+(\bcol{x}) \,\right) \, V^+(\bcol{x})
			\\
			=&\, \left| V^\ast(x) - V(x)\right|
			\\
			\kcol{\geq}&\, V(x) - V^\ast(x)
		\end{aligned}
	\end{equation}
	where the inequalities hold for all $\xinX$. The first inequality comes from the definition of the weighted $\infty$-norm. The first equality holds as $V^+$ is a strictly positive function, and the final inequality stems from the definition of $|\cdot|$.
	
	Thus,
	\begin{equation} \nonumber
		\begin{aligned}
			&\,V(x) - (\mcal{T}^M V)(x)
			\\
			\leq&\, \bcol{ \epsilon \, V^+(x) + V^\ast(x) } - (\mcal{T}^M V)(x)
			\\
			\leq&\, \epsilon \, V^+(x) + \left|\, (\bcol{\mcal{T}^M} V^\ast)(x) - (\mcal{T}^M V)(x) \,\right|
			\\
			\leq&\, \epsilon \, V^+(x) + \bcol{ \gamma^M \left( H_V^M \left(\, \left| \kcol{ V^\ast - V } \right| \,\right) \right) }(x)
			\\
			\leq&\, \epsilon \, V^+(x) + \gamma^M \left( H_V^M \left(\, \bcol{ \epsilon \, V^+ } \,\right) \right)(x)
			\\
			=&\, \epsilon \, V^+(x) + \gamma^M \, \bcol{\epsilon} \, \left( H_V^M  V^+ \right)(x)
			\\
			=&\, \left(\, V^+(x) + \gamma^M \, \left( H_V^M  V^+ \right)(x) \,\right) \, \epsilon
		\end{aligned}
	\end{equation} 
	where the inequalities hold for all $\xinX$. The first inequality is a consequence of \eqref{eq:01_for_T_iterated_V_geq_V_minusLyap}. The second inequality uses the fact that $V^\ast \!=\! \mcal{T}^M V^\ast$ and the definition of $|\cdot|$. The third inequality is a direct application of Lemma \ref{lemma:T_iterated_V_diff_leq_H_iterated_V_diff}. The fourth inequality uses \eqref{eq:01_for_T_iterated_V_geq_V_minusLyap} and the monotonicity of operator $H_V^M$. The two equalities follow from simple algebra. 
\end{IEEEproof}

\vspace{0.2cm}

\begin{lemma} \label{lemma:lyapunov_function_bound_V_HV}
	Given any Lyapunov function $V$ (Definition \ref{def:lyap_func_Vform}), and its respective Lyapunov constant $\beta_{V}$, then,
	\begin{equation} \nonumber
		\begin{aligned}
			&\left( \frac{2}{1-\beta^{M}_{V} } - 1 \right) \,
				\left( V(x)\,-\, \gamma^M \, (H_V^M V)(x) \, \right)
			\\
			&\qquad\qquad\qquad \geq\,	
				\left( V(x) \,+\, \gamma^M (H_V^M V)(x) \right)
		\end{aligned}
	\end{equation}
	for all $\xinX$.
\end{lemma}

\vspace{0.2cm}

\begin{IEEEproof}
	By the definition of the Lyapunov function that $(HV)(x) \leq (\beta_V/\gamma) \, V(x)$ for all $\xinX$, thus we get that,
	\begin{equation} \nonumber
		\begin{aligned}
			 \left(\, H_V^{M} \, V \,\right)(x)
				\,=&\, \left(\, H_V^{M-1} \, (H_V V) \,\right)(x)
			\\
			\kcol{\leq}&\, \left(\, H_V^{M-1} \, \left( (\beta_{V}/\gamma) V \right) \,\right)(x)
			\\
			=&\, (\beta_{V}/\gamma) \, \left(\, H_V^{M-1} \, V \,\right)(x)
		\end{aligned}
	\end{equation}
	where the inequality holds for all $\xinX$ by the monotone property of $H^k$ for any $k\in\mbb{N}$. Iterating the same argumentation $M$-times leads to,
	\begin{equation} \nonumber
		\begin{aligned}
			\left(\, H_V^{M} \, V \,\right)(x)
				\,\leq\, \left(\beta_{V}/\gamma\right)^M \, V(x)
				\,,
		\end{aligned}
	\end{equation}
	for all $\xinX$. As $V$ is strictly positive, this implies that,
	\begin{equation} \nonumber
		\begin{aligned}
			&\, \frac{2}{1-\frac{\gamma^M \, (H_V^M V)(x)}{V(x)}} \,-\, 1 \,\leq\, \frac{2}{1-\beta_V^{M}} \,-\, 1
				\,,
		\end{aligned}
	\end{equation}
	for all $\xinX$. Manipulating the left-hand-side,
	\begin{equation} \nonumber
		\begin{aligned}
			&\, \left( \frac{2}{ 1 \,-\, \frac{\gamma^M \, (H_V^M V)(x)}{V(x)} } - 1 \right)
				\,=\,
				\frac{V(x) \,+\, \gamma^M \, (H_V^M V)(x)}{ V(x) \,-\, \gamma^M \, (H_V^M V)(x) }
				\,.
		\end{aligned}
	\end{equation}
	Hence the result follows.
\end{IEEEproof}

\vspace{0.2cm}

\begin{lemma} \label{lemma:feasible_forIterated_V_by_shifting}
	Let $\hat{V}^{+}(x)$ be a Lyapunov function (Definition \ref{def:lyap_func_Vform}) and $\hat{V} \in \approxFuncSpaceX$ arbitrary, and define $\tilde{V}$ as,
	\begin{equation} \label{lemma:feasible_forIterated_V_by_shifting_eq01}
		\tilde{V}(x) = \hat{V}(x)
			\,-\,
			\epsilon \,
			\left( \frac{2}{1-\beta^{M}_{\hat{V}^{+}} } - 1 \right)
			\, \hat{V}^{+}(x)
	\end{equation}
	where $\epsilon = \| V^\ast - \hat{V} \|_{\infty,1/\hat{V}^{+}}$, then $\tilde{V}(x) \leq \left( \mcal{T}^M \tilde{V} \right)(x)$ for all $\xinX$, i.e., it is feasible for the approximate iterated LP. Additionally, if $\hat{V}^{+} \in \approxFuncSpaceX$ then $\tilde{V}$ is an element of $\approxFuncSpaceX$.
\end{lemma}

\vspace{0.2cm}

\begin{IEEEproof}
	Starting from the right-hand-side of the iterated Bellman inequality,
	\begin{equation} \nonumber
		\begin{aligned}
			&\, (\mcal{T}^M \tilde{V})(x)
			\\
			=&\, \bcol{ (\mcal{T}^M \hat{V})(x)
				- (\mcal{T}^M \hat{V})(x) } + (\mcal{T}^M \tilde{V})(x)
			\\
			\geq&\, (\mcal{T}^M \hat{V})(x)
				- \bcol{ \left|\, \kcol{ (\mcal{T}^M \hat{V})(x) - (\mcal{T}^M \tilde{V})(x) } \,\right| }
			\\
			\geq&\, (\mcal{T}^M \hat{V})(x)
				- \bcol{\gamma^M \, \left(\, H_V^M \, \left| \kcol{\hat{V}(x) \,-\, \tilde{V}(x) } \right| \,\right) }
			\\
			=&\,  (\mcal{T}^M \hat{V})(x)
				- \gamma^M \, \epsilon \, \left( \frac{2}{1-\beta^{M}_{\hat{V}^{+}} } - 1 \right) \, \big(\, H^M \, \hat{V}^{+} \,\big) \, (x)
			\\
			\kcol{\geq}&\, \bcol{ \hat{V}(x) \,-\, \epsilon \, \left( \hat{V}^{+}(x) \,+\, \gamma^M \, \big(H^M \hat{V}^{+}\big)(x) \right) }
				\\
				&\qquad-\, \gamma^M \, \epsilon \, \left( \frac{2}{1-\beta^{M}_{\hat{V}^{+}} } - 1 \right)  \, \big(\, H^M \hat{V}^{+} \,\big)(x)
			\\
		\end{aligned}
	\end{equation}
	\begin{equation} \nonumber
		\begin{aligned}
			=&\, \bcol{ \tilde{V}(x) } \,-\, \epsilon \, \left(\, \hat{V}^{+}(x) \,+\, \gamma^M \, \big( H^M \hat{V}^{+} \big)(x) \,\right)
				\\
				&\quad +\, \kcol{\epsilon \left( \frac{2}{1-\beta^{M}_{\hat{V}^{+}} } - 1 \right)}
				\, \Bigg( \bcol{ \hat{V}^{+}(x) } - \gamma^M \left( H^M \hat{V}^{+} \right)(x) \Bigg)
			\\
			\geq&\, \tilde{V}(x)
		\end{aligned}
	\end{equation}
	where the inequality holds for all $\xinX$. The first equality is simple algebra and the first inequality is from the definition of $|\cdot|$. The second inequality is a direct application of Lemma \ref{lemma:T_iterated_V_diff_leq_H_iterated_V_diff}. The second equality follows from the definition of $\tilde{V}$ given in \eqref{lemma:feasible_forIterated_V_by_shifting_eq01}. The third inequality stems from applying Lemma \ref{lemma:T_iterated_V_geq_V_minusLyap} to the $(\mcal{T}^M \hat{V})$ term. The last equality again uses the definition of $\tilde{V}$ and the last inequality follows from Lemma \ref{lemma:lyapunov_function_bound_V_HV}.
	
	By \eqref{lemma:feasible_forIterated_V_by_shifting_eq01}, $\tilde{V}$ is a linear combination of $\hat{V}$ and $\hat{V}^+$. As $\hat{V}$ and $\hat{V}^+$ are both elements of $\approxFuncSpaceX$, so is $\tilde{V}$.
\end{IEEEproof}

\vspace{0.2cm}

\begin{IEEEproof}[Proof of of Theorem \ref{theorem:bound_lyapunov_iterated_Vform}]

	Given any $\hat{V}(x)\in \approxFuncSpaceX$, construct $\tilde{V} \in \approxFuncSpaceX$ following Lemma \ref{lemma:feasible_forIterated_V_by_shifting} to be feasible for the approximate iterated LP. Working from the left hand side of the bound,
	\begin{equation} \nonumber
		\begin{aligned}
			&\, \left\|\, V^\ast \,-\, \hat{V}^\ast \,\right\|_{1,c}
			\\
			\kcol{\leq}&\, \left\|\, V^\ast \,-\, \bcol{\tilde{V}} \,\right\|_{1,c}
			\\
			=&\, \int_{\mcal{X}} \, \left(\frac{\hat{V}^{+}(x)}{\hat{V}^{+}(x)}\right) \, \left|\, V^\ast(x) \,-\, \tilde{V}(x) \,\right| \, c(\intd{x})
			\\
			\kcol{\leq}&\, \left(\, \int_{\mcal{X}} \, \hat{V}^{+}(x) \, c(\intd{x}) \,\right) \,\, \bcol{ \sup_{z\in\mcal{X}} \frac{ \left|\, V^\ast(z) \,-\, \tilde{V}(z) \,\right|}{\hat{V}^{+}(z)} }
			\\
			=&\, \left(\, \bcol{ \left\| \hat{V}^{+} \right\|_{1,c(x)} } \,\right) \,\, \bcol{ \left\|\, V^\ast \,-\, \tilde{V} \,\right\|_{\infty,1/\hat{V}^{+} } }
			\\
			\kcol{\leq}&\, \left\| \hat{V}^{+} \right\|_{1,c(x)}
			\left( \left\| V^\ast - \bcol{ \hat{V} } \right\|_{\infty,1/\hat{V}^{+}} + \left\| \bcol{ \hat{V} } - \tilde{V} \right\|_{\infty,1/\hat{V}^{+} } \right)
			\\
			=&\, \left\| \hat{V}^{+} \right\|_{1,c(x)} \, \left( \frac{2}{1-\beta^{M}_{\hat{V}^{+}} } \right) \, \left\| V^\ast - \hat{V} \right\|_{\infty,1/\hat{V}^{+}} 
		\end{aligned}
	\end{equation}
	where the inequalities hold for all $\xinX$. The first inequality follows from Lemma \ref{lemma:approxLP_for_Q_solves_min_1norm} and Lemma \ref{lemma:feasible_forIterated_V_by_shifting}. The first equality is the definition of the weighted $1$-norm and holds as $\hat{V}^+$ is strictly positive. The second inequality holds because the objective of the supremum is non-negative for all $z\!\in\!\mcal{X}$. The second equality is the definition of the weighted $1$-norm and weighted $\infty$-norm. The final inequality follows by the triangle inequality. The final equality stems from using \eqref{lemma:feasible_forIterated_V_by_shifting_eq01} by taking the weighted $\infty$-norm of $(\hat{V}-\tilde{V})$ and then some simple algebra.
	As the inequality established holds for any $\hat{V}(x)\in\approxFuncSpaceX$, it also holds when the infimum over all $\hat{V}(x)\in\approxFuncSpaceX$ is taken on the right-hand-side.
	
\end{IEEEproof}

%% file: sec/11_appendix_F_unify.tex


\section{Proofs of equivalent \textQ-function formulation} \label{app:Qform_equivalence}

\begin{IEEEproof}[Proof of of Lemma \ref{lemma:Qform_Vform_equivalence}]
	
	We shall show that any feasible solution of \eqref{eq:PropOfEquiv_forQ} corresponds to a feasible solution of \eqref{eq:PropOfEquiv_forV} with the same objective value, and vice versa.
	Note that for the proof superscript $(\cdot)^\prime$ indicates a decision variable of problem \eqref{eq:PropOfEquiv_forV}.
	
	Suppose that $\{\hat{Q}_j\}_{j=0}^{M-1}$ , $\{\hat{V}_j\}_{j=0}^{M-1}$ is a feasible solution of \eqref{eq:PropOfEquiv_forQ}, and take the following decision variables for \eqref{eq:PropOfEquiv_forV},
		\begin{equation} \nonumber
			\begin{aligned}
				\hat{Q}_{0}^\prime = \hat{Q}_{0},
					\qquad \hat{V}_{j}^\prime = \hat{V}_{j},\, j=0,\dots,M-1
					\,.
			\end{aligned}
		\end{equation}
	We now check feasibility for the constraints of \eqref{eq:PropOfEquiv_forV}.
		\begin{equation} \nonumber
			\begin{aligned}
				\hat{Q}_{0}^\prime(x,u)
					= \hat{Q}_0(x,u)
					\leq \mcal{T}_u \hat{V}_{0}(x,u)
					= \mcal{T}_u \hat{V}^\prime_{0}(x,u)
					\,,
			\end{aligned}
		\end{equation}
	for all $\smash{\xinX}$ and $\smash{\uinU}$, thus \eqref{eq:PropOfEquiv_forV_01} is satisfied. We have that for $\smash{j=1,\dots,M\!-\!1}$,
		\begin{equation} \nonumber
			\begin{aligned}
					\hat{V}_{j-1}^\prime(x)
						= \hat{V}_{j-1}(x)
						\leq \hat{Q}_{j}(x,u)
						\leq \mcal{T}_u \hat{V}_{j}(x,u)
						= \mcal{T}_u \hat{V}_{j}^\prime(x,u)
						\,,
			\end{aligned}
		\end{equation}
	for all $\smash{\xinX}$ and $\smash{\uinU}$, thus \eqref{eq:PropOfEquiv_forV_02} are satisfied. Finally,
		\begin{equation} \nonumber
			\begin{aligned}
				\hat{V}_{M-1}^\prime(x)
					= \hat{V}_{M-1}(x)
					\leq \hat{Q}_{0}(x,u)
					= \hat{Q}_{0}^\prime(x,u)
					\,,
			\end{aligned}
		\end{equation}
	for all $\smash{\xinX}$ and $\smash{\uinU}$, thus \eqref{eq:PropOfEquiv_forV_03} is also satisfied, and the considered decision variables are feasible for problem \eqref{eq:PropOfEquiv_forV}. As $\hat{Q}_{0}^\prime = \hat{Q}_{0}$, the objective values are equal. This completes the equivalence in one direction.
	
	Suppose that $\hat{Q}_{0}^\prime$, $\{\hat{V}_{j}^\prime\}_{j=0}^{M-1}$ is a feasible solution of \eqref{eq:PropOfEquiv_forV}, and take the following decision variables for \eqref{eq:PropOfEquiv_forQ},
		\begin{equation} \nonumber
			\begin{aligned}
				\hat{Q}_{0} =&\, \hat{Q}_{0}^\prime,
				\\
				\hat{V}_{j} =&\, \hat{V}_{j}^\prime
					\,,\,\, &&j=0,\dots,M-1
					\,,
				\\
				\hat{Q}_{j} =&\, \mcal{T}_u \hat{V}_{j}^\prime
					\,,\,\, &&j=1,\dots,M-1	
					\,,
			\end{aligned}
		\end{equation}
	where the choices of $\hat{Q}_j$ are valid by the assumption.
	We now check the feasibility for the constraints of \eqref{eq:PropOfEquiv_forQ}.
		\begin{equation} \nonumber
			\begin{aligned}
				\hat{Q}_{0}(x,u)
					= \hat{Q}_{0}^\prime(x,u)
					\leq \mcal{T}_u \hat{V}_{0}^\prime(x,u)
					= \mcal{T}_u \hat{V}_{0}(x,u)
					\,,
			\end{aligned}
		\end{equation}
	for all $\smash{\xinX}$ and $\smash{\uinU}$, and for $\smash{j=1,\dots,M\!-\!1}$  we have,
		\begin{equation} \nonumber
			\begin{aligned}
				\hat{Q}_{j}(x,u)
					= \mcal{T}_u \hat{V}_{j}^\prime(x,u)
					\leq \mcal{T}_u \hat{V}_{j}(x,u)
					\,,
			\end{aligned}
		\end{equation}
	for all $\smash{\xinX}$ and $\smash{\uinU}$, thus \eqref{eq:PropOfEquiv_forQ_01} are satisfied. We have that for $\smash{j=0,\dots,M\!-\!2}$,
		\begin{equation} \nonumber
			\begin{aligned}
				\hat{V}_{j}(x)
					= \hat{V}_{j}^\prime(x)
					\leq \mcal{T}_u \hat{V}_{j+1}^\prime(x,u)
					= \hat{Q}_{j+1}(x,u)
					\,,
			\end{aligned}
		\end{equation}
	for all $\smash{\xinX}$ and $\smash{\uinU}$, thus \eqref{eq:PropOfEquiv_forQ_02} are also satisfied. Finally,
		\begin{equation} \nonumber
			\begin{aligned}
				\hat{V}_{M-1}(x)
					= \hat{V}_{M-1}^\prime(x)
					\leq \hat{Q}_{0}^\prime(x,u)
					= \hat{Q}_0(x,u)
					\,,
			\end{aligned}
		\end{equation}
	for all $\smash{\xinX}$ and $\smash{\uinU}$, thus \eqref{eq:PropOfEquiv_forQ_03} is also satisfied, and the considered decision variables are feasible for problem \eqref{eq:PropOfEquiv_forQ}. As $\hat{Q}_{0} = \hat{Q}_{0}^\prime$, the objective values are equal.
\end{IEEEproof}

%% file: sec/11_appendix_G_lyap_computations.tex

\section{Computing $V^\ast$, $\hat{V}$, $\mu$, and Lyapunov Functions} \label{app:supplement_for_performance_bounds_numerical}

This appendix provides additional details for the numerical example of Section \ref{sec:numerical_1d}.

\begin{figure}
	\iftoggle{doublecolumn}{
		\begin{center}
			\includegraphics[width = 0.40 \textwidth]{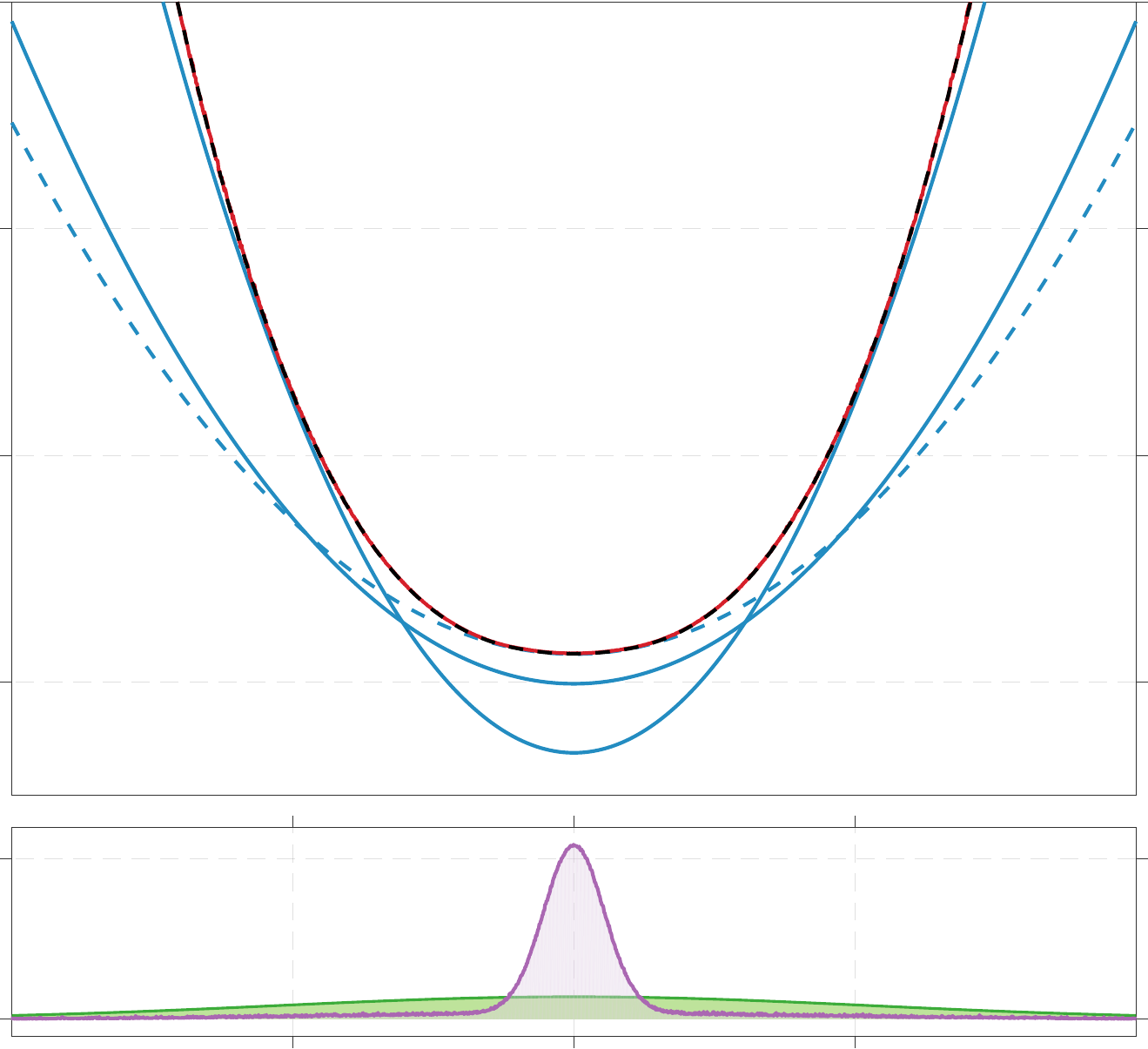}
		\end{center}
		\vspace{0.20cm}
		%
	}{
		\begin{center}
			\includegraphics[width = 7.1cm]{img/Sys001D_005_Vhat_and_mu_v02.pdf}
		\end{center}
		\vspace{-0.28cm}
		\hspace{3.57cm}
		%
	}
	%
	\begin{tikzpicture}[overlay]
		\node[align=center , rotate=0] at (4.4cm,0.1cm) { State Space, $\mcal{X}$ };
		\node[align=center , rotate=0] at (2.60cm,0.60cm) { $-3.0$ };
		\node[align=center , rotate=0] at (4.48cm,0.60cm) { $0.0$ };
		\node[align=center , rotate=0] at (6.24cm,0.60cm) { $3.0$ };
		\node[align=right , rotate=0] at (0.55cm,1.02cm) { $0.0$ };
		\node[align=right , rotate=0] at (0.55cm,2.00cm) { $0.9$ };
		\node[align=right , rotate=0] at (0.60cm,3.10cm) { $0$ };
		\node[align=right , rotate=0] at (0.60cm,4.50cm) { $20$ };
		\node[align=right , rotate=0] at (0.60cm,5.90cm) { $40$ };
		\draw[gray70,line width = 0.5pt,fill=white] (3.3cm,5.6cm) rectangle (5.3cm,7.0cm);
		\draw[black,line width = 1.2pt,dashed] (3.5cm,6.7cm) -- (4.1cm,6.7cm);
		\node[right] at (4.3cm,6.7cm) {{\small$V^\ast$}};
		\draw[myblue,line width = 1.2pt,solid] (3.5cm,6.3cm) -- (4.1cm,6.3cm);
		\node[right] at (4.3cm,6.3cm) {\color{myblue}{\small$\hat{V}^\ast$}};
		\draw[myred ,line width = 1.2pt,solid] (3.5cm,5.9cm) -- (4.1cm,5.9cm);
		\node[right] at (4.3cm,5.9cm) {\color{myred}{\small$V_{\hat{\pi}}$}};
		\draw[gray70,line width = 0.5pt,fill=white] (5.2cm,1.5cm) rectangle (7.7cm,2.1cm);
		\draw[mygreen,line width = 0.8pt,fill=myltgreen] (5.4cm,1.7cm) rectangle (5.8cm,1.9cm);
		\node[right] at (5.9cm,1.8cm) {\color{mygreen}{\small$\nu$}};
		\draw[mypurple,line width = 0.8pt,fill=myltpurple] (6.6cm,1.7cm) rectangle (7.0cm,1.9cm);
		\node[right] at (7.1cm,1.8cm) {\color{mypurple}{\small$\tilde{\mu}$}};
		\node[align=center, myblue, rotate=00] at (4.40cm,2.96cm) {{\small$M=1$}};
		\node[align=center, myblue, rotate=00] at (4.40cm,2.54cm) {{\small$M=200$}};
		\node[align=center, myblue, rotate=65] at (7.45cm,6.60cm) {{\small$M=1$}};
		\node[align=center, myblue, rotate=75] at (6.94cm,6.60cm) {{\small$M=200$}};
		\node[align=center, myblue, rotate=57] at (7.50cm,5.50cm) {{\small LQR}};
	\end{tikzpicture}
	\caption[Value functions and occupancy measures]
	{
		Value functions and occupancy measures for the 1-dimensional example of Section \ref{sec:numerical_1d}. The approximate value functions $\hat{V}^\ast$ (blue) are labelled with the number of $M$ bellman iterations used, and are point-wise under-estimators of $V^\ast$ (dashed black). The online performance $V_{\hat{\pi}}$ (red) and discounted occupancy measure $\tilde{\mu}$ (purple) are shown for the approximate policy arising from $M\!=\!200$, for the $\smash{M\!=\!1}$ and LQR policies the $V_{\hat{\pi}}$ and $\tilde{\mu}$ results are indistinguishable on the scale of this graph. The initial state distribution $\nu$ (green) is shown for comparison, and $c(\cdot) \!=\! \nu(\cdot)$ was used for computing $\hat{V}^\ast$. The dashed blue line labelled LQR is the approximate value functions that arises from using $\spaceU \in \rdim{}$, i.e., unconstrained input, in the approximate LP.
	}
	\label{fig:1d_example_Vfunctions}
\end{figure}

%
The value function was computed on the interval $\smash{\spaceX \!=\! [-12\sigma_{\nu},12\sigma_{\nu}]}$ at $10^4$ evenly spaced discretization points.
%
The $V_{\hat{\pi}}$ and $\tilde{\mu}$ are computed for $10^4$ $x_0$ samples from $\nu$. The expectation with respect to $\xi$ is empirically evaluated using $10^4$ extractions from the disturbance process, different for each $x_0$, and each is simulated for $10^3$ time steps.
%
The boundary of $\smash{\spaceX \!=\! [-12\sigma_{\nu},12\sigma_{\nu}]}$ was not reached by any sample.
%
Fig. \ref{fig:1d_example_Vfunctions} shows on the upper plot $V^\ast$ (black dashed), the approximate value functions, $\hat{V}^\ast$ (blue), and the online performance, $V_{\hat{\pi}}$ (red). The lower plot depicts the initial state distribution $\nu$ (green), and the discounted state occupancy measure $\tilde{\mu}$ (purple) that arises from playing the approximate policy.
%
The $V_{\hat{\pi}}$ and $\tilde{\mu}$ are shown only for $\smash{M\!=\!200}$ Bellman iterations because they are similar for all choices of $M$.

%
Fig. \ref{fig:1d_example_Vfunctions} provides the visual insight necessary to explain the numerical trend observed in the data of Table \ref{tab:OnlineBounds_for1DExample} that for $\smash{M\!=\!200}$ the online performance of the greedy policy is slightly worse and the bound significantly more conservative.
It is clear from Fig. \ref{fig:1d_example_Vfunctions} that $\hat{V}^\ast$ with $\smash{M\!=\!1}$ gives a better point-wise lower-bound in the region near $\smash{x\!=\!0}$, compared to $M\!=\!200$. As $\tilde{\mu}$ is more concentrated near $\smash{x\!=\!0}$ than $\nu$, the bound is tighter.
%
The difference in online performance is also explained by the difference of the approximate value functions in the region near $\smash{x\!=\!0}$. As the greedy policy is closely related to the gradient of the value function, in regions where the gradient of a $\hat{V}$ closely approximates that of $V^\ast$, the approximate greedy policy will generate near-optimal control actions.
In Fig. \ref{fig:1d_example_Vfunctions} it is clear that in the region near the origin $\hat{V}^\ast_{M=1}$ matches the gradient of $V^\ast$ much better than $\hat{V}^\ast_{M=200}$.
Due to the input constraints of this problem, outside of that region all value functions that rise steeply enough lead to the same performance because the input saturates at $\pm 1$.

To explain the computation of Lyapunov functions, first recall that the restricted function space used for the one dimensional example is the space of univariate quadratics, with $\smash{p\in\rdim{}}$ as the quadratic coefficient, $\smash{s\in\rdim{}}$ as the constant offset, and the linear term omitted.
From the definition of $\beta_V$ and the $H_V$ operator, it is clear that if a function $V$ is a Lyapunov function then $\smash{\alpha V \,:\, x\,\mapsto \alpha \, V(x)}$, with $\smash{\alpha \!\in\! \rdim{}_{++}}$, is also a Lyapunov function. Moreover, the right-hand-side of Theorem \ref{theorem:bound_lyapunov_iterated_Vform} is unchanged by this positive scaling. Thus, without loss of generality we fix $\smash{s\!=\!1}$ and parametrize candidate Lyapunov functions by the quadratic co-efficient.

\begin{figure}
	\iftoggle{doublecolumn}{
		\begin{center}
			\includegraphics[width = 6.0cm]{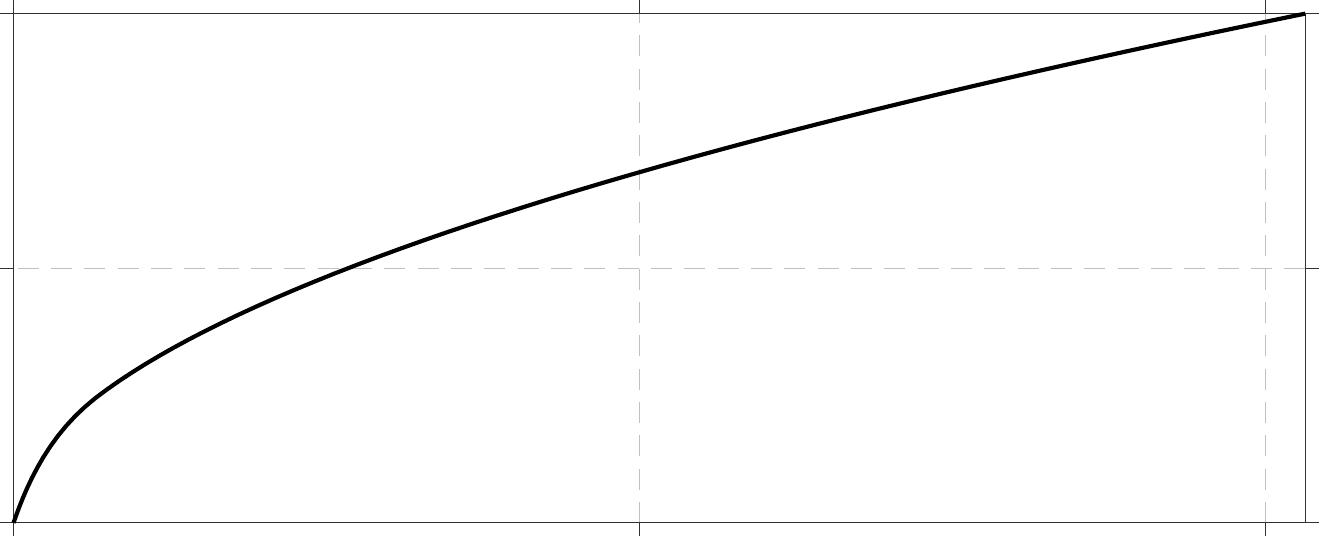}
		\end{center}
		\vspace{0.10cm}
	%
	}{
		\begin{center}
			\includegraphics[width = 6.1cm]{img/lyap_beta_vs_P.pdf}
		\end{center}
		\vspace{-0.35cm}
		\hspace{3.47cm}
	%
	}
	%
	\begin{tikzpicture}[overlay]
	\node[align=center , anchor=south  , rotate=0] at (4.3cm,-0.15cm) {\small{Quadratic coefficient, $p$}};
	\node[align=center , anchor=center , rotate=90] at (0.3cm,1.9cm) {\small{ $\beta_V$ }};
	\node[align=center , anchor=north , rotate=0] at (1.55cm,0.75cm) {\small{ $0.0$ }};
	\node[align=center , anchor=north , rotate=0] at (4.4cm,0.75cm) {\small{ $0.005$ }};
	\node[align=center , anchor=north , rotate=0] at (7.3cm,0.75cm) {\small{ $0.01$ }};
	\node[align=right , anchor=east , rotate=0] at (1.60cm,0.86cm) {\small{ $0.95$ }};
	\node[align=right , anchor=east , rotate=0] at (1.60cm,3.08cm) {\small{ $1.00$ }};
	\end{tikzpicture}
	\caption[Lyapunov function $\beta_V$ versus $P$]
	{
		Set of Lyapunov functions, parametrized by the quadratic coefficient, for the one dimensional example of Section \ref{sec:numerical_1d} and the corresponding $\beta_V$.
	}
	\label{fig:appendix:lyap_beta_vs_P}
\end{figure}

To compute the set of Lyapunov functions and their corresponding $\beta_V$ value, we take a brute force approach. As discussed in Section \ref{sec:bounds_fitting_lyap}, a constant function, i.e., $\smash{p\,=\,0}$, is a Lyapunov function with $\smash{\beta_V \,=\, \gamma}$. For this system, with stable linear dynamics, $\beta_V$ increases with $p$. To find the set of Lyapunov functions, we increase $\smash{p \,>\, 0}$ in small increments, and compute the value of $\beta_V$ by discretizing the state space on a sufficiently large interval. The relationship of $\beta_V$ versus $p$ is shown in Fig. \ref{fig:appendix:lyap_beta_vs_P}.

\begin{figure}
	\iftoggle{doublecolumn}{
		\begin{center}
			\includegraphics[width = 6.0cm]{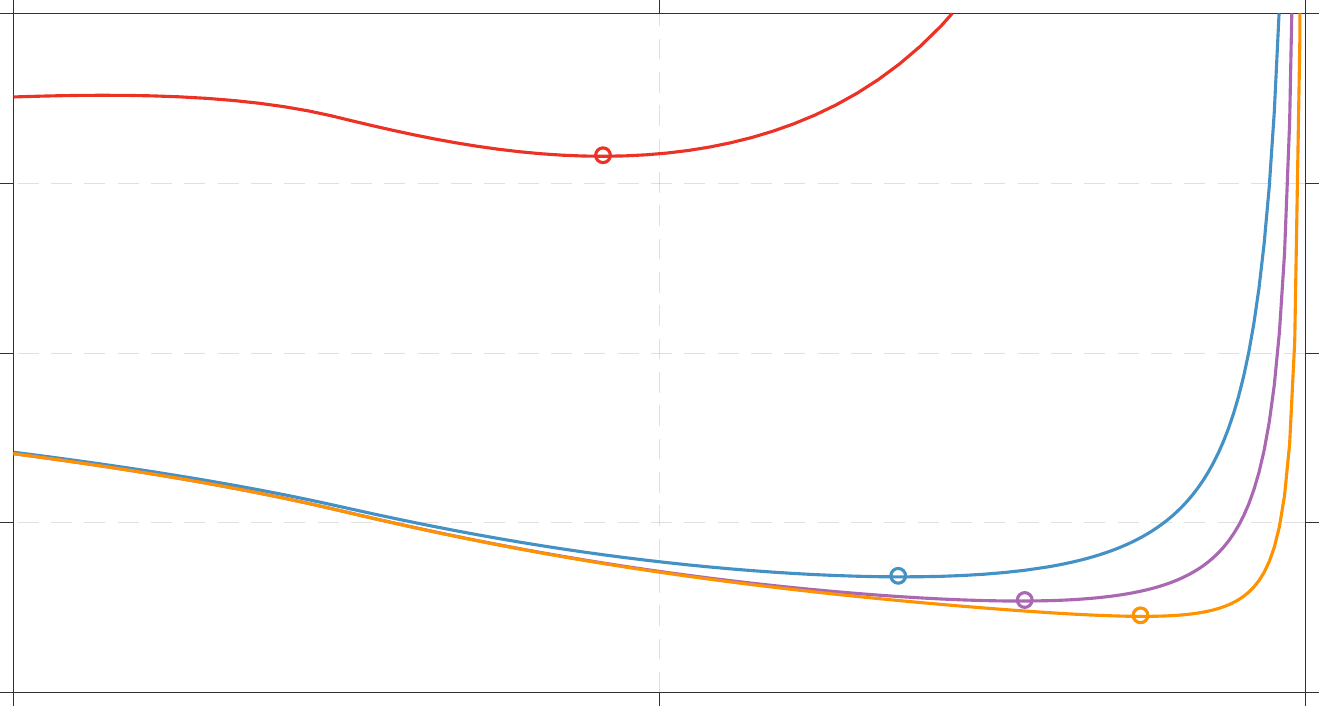}
		\end{center}
		\vspace{0.10cm}
	}{
		\begin{center}
			\includegraphics[width = 6.00cm]{img/lyap_RHS_vs_beta.pdf}
		\end{center}
		\vspace{-0.38cm}
		\hspace{3.53cm}
	%
	}
	%
	\begin{tikzpicture}[overlay]
	\node[align=center , anchor=south  , rotate=0] at (4.3cm,-0.15cm) {\small{ $\beta_V$ }};
	\node[align=center , anchor=center , rotate=90] at (0.3cm,2.4cm) {\small{RHS of Theorem \ref{theorem:bound_lyapunov_iterated_Vform}}};
	\node[align=center , anchor=north , rotate=0] at (1.55cm,0.75cm) {\small{ $0.95$ }};
	\node[align=center , anchor=north , rotate=0] at (4.44cm,0.75cm) {\small{ $0.975$ }};
	\node[align=center , anchor=north , rotate=0] at (7.40cm,0.75cm) {\small{ $1.0$ }};
	\node[align=right , anchor=east , rotate=0] at (1.60cm,0.86cm) {\small{ $0$ }};
	\node[align=right , anchor=east , rotate=0] at (1.60cm,2.34cm) {\small{ $2000$ }};
	\node[align=right , anchor=east , rotate=0] at (1.60cm,3.84cm) {\small{ $4000$ }};
	\node[align=left , anchor=west , rotate=90] at (4.17cm,1.70cm) {\small{$M\!=\!10$}};
	\node[align=left , anchor=west , rotate=90] at (5.51cm,1.70cm) {\small{$M\!=\!100$}};
	\node[align=left , anchor=west , rotate=90] at (6.09cm,1.70cm) {\small{$M\!=\!200$}};
	\node[align=left , anchor=west , rotate=90] at (6.62cm,1.70cm) {\small{$M\!=\!500$}};
	\draw[matlabred    ,line width = 0.5pt] (4.175,2.85) -- (4.175,3.20);
	\draw[matlabblue   ,line width = 0.5pt] (5.515,1.80) -- (5.515,1.36);
	\draw[matlabpurple ,line width = 0.5pt] (6.090,1.80) -- (6.090,1.24);
	\draw[matlaborange ,line width = 0.5pt] (6.620,1.80) -- (6.620,1.17);
	\end{tikzpicture}
	\caption[Lyapunov function $\beta_V$ versus $P$]{
		Right-hand-side of Theorem \ref{theorem:bound_lyapunov_iterated_Vform} for the one dimensional example of Section \ref{sec:numerical_1d}, evaluated for the choice $\smash{c(\cdot) \!=\! \nu(\cdot)}$. The circles mark the minimum for each curve, showing that for each $M$, a different Lyapunov function achieves the tightest bound.
	}
	\label{fig:appendix:lyap_RHS_vs_beta}
\end{figure}

All Lyapunov functions yield a valid bound, and the Lyapunov with the tightest bound changes based on the number of Bellman inequality iterations $M$. To provide some insight, Fig. \ref{fig:appendix:lyap_RHS_vs_beta} shows the right-hand-side of Theorem \ref{theorem:bound_lyapunov_iterated_Vform} for the choice $\smash{c(\cdot) \!=\! \nu(\cdot)}$ versus $\beta_V$ for this example.

%% file: sec/11_appendix_H_SprocedureReformulation.tex

\section{Reformulation of Bellman Inequality} \label{app:Sprocedure_reformulation}

This appendix provides a sufficient reformulation of the Bellman Inequality that is used in the numerical examples for solving programs \eqref{eq:LP_approach_to_ADP_iterated_Vform} and \eqref{eq:PropOfEquiv_forQ} to find an approximate value function and \textQ-function respectively.
%
See Section \ref{sec:numerical_nd} for the definitions of $A$, $B_u$, and $B_\xi$ as the linear dynamics, and \eqref{eq:quadratic_basis_function_space} for the specification of the quadratic basis functions.
%
We introduce $\smash{\underline{u}_i,\overline{u}_i \in \mbb{R}}$, $\smash{i\!=\!1,\dots,n_u}$, to denote the lower and upper bounds that describe each coordinate of the $\smash{\mcal{U} \subseteq \mbb{R}^{n_u}}$ space.
%
To concisely represent the quadratic stage cost we introduce the matrix $\smash{L \in \mbb{R}^{(n_x+n_u+1)\times(n_x+n_u+1)}}$ that takes the the form $\smash{l(x,u) = [x^\tran,u^\tran,1] \, L \, [x^\tran,u^\tran,1]^\tran}$.
%
The notation $\smash{\diag{\cdot}}$ places the vector argument on the diagonal of an otherwise zero matrix, and $e_i$ is the standard basis column vector with $1$ in the $i^{\mrm{th}}$ element and zeros elsewhere, with the dimension clear from context.

Using this notation, each inequality of the form $\smash{\hat{Q}_{j}(x,u) \,\leq\, \mcal{T}_u \hat{V}_{j}(x,u)}$ for all $\smash{x\in\mbb{R}^{n_x}}$, $\smash{\uinU}$ is sufficiently reformulated as the following LMI:
	\begin{equation} \nonumber
		\begin{aligned}
			0 \,\preceq&\,
				-\,
				\begin{bmatrix}
					P_j^Q & \frac{1}{2} p_j^Q \\ \star & s_j^Q
				\end{bmatrix}
				\,+\,
				L
			\\
			&\,+\, \gamma \; 
				\begin{bmatrix}
					A^\tran P_j A
						& A^\tran P_j B_u
						& \frac{1}{2} A^\tran p_j + A^\tran P_j B_\xi \expval{}{\xi}
					\\
					\star
						& B_u^\tran P_j B_u
						& \frac{1}{2} B_u^\tran p_j + B_u^\tran P_j B_\xi \expval{}{\xi}
					\\
					\star
						& \star
						& s_j + \trace{ B_\xi^\tran P_j B_\xi \expval{}{\xi \xi^\tran}}
				\end{bmatrix}
			\\
			&\,-\, \sum\limits_{i=1}^{n_u} \, \lambda_i \,
				\begin{bmatrix}
					0_{n_x \times n_x} & 0 & 0
					\\
					\star & -\smash{\diag{e_i}}
						& \smash{\frac{1}{2}\left( \underline{u}_i + \overline{u}_i \right)} e_i
					\\
					\star
						& \star
						& -\underline{u}_i \, \overline{u}_i
				\end{bmatrix}
				\,,
		\end{aligned}
	\end{equation}
%
where $\star$ indicates that the matrix is symmetric, and the $\smash{\lambda_i \in \mbb{R}_+}$, $\smash{i\!=\!1,\dots,n_u}$, are the auxiliary variables introduced when using the S-procedure to reformulate the for all $\smash{\uinU}$ part of the constraint.

The objective function of programs \eqref{eq:LP_approach_to_ADP_iterated_Vform} and \eqref{eq:PropOfEquiv_forQ} is linear in the decision variables and evaluation of the objective requires the first and second moments of the relevance weighting parameter. For the \textQ-function formulation, the objective is:
	\begin{equation} \nonumber
		\max_{P_0^Q,p_0^Q,s_0^Q} \hspace{0.2cm} \trace{{P_0^Q \, \Sigma_c}} \,+\, \mu_c^\tran \, p_0^Q \,+\, s_0^Q
			\,,
	\end{equation}
where $\smash{\mu_c \in \rdim{n_x+n_u}}$ and $\smash{\Sigma_c \in \mbb{S}^{n_x+n_u}}$ are the first and second moments of the measure $c(x,u)$, and $\trace{\cdot}$ denotes the trace of a square matrix.